\RequirePackage{fix-cm}
\documentclass[smallextended]{svjour3}       
\smartqed  
\usepackage{graphicx}
\usepackage{multirow}
\usepackage[margin=0.5in]{geometry}
\usepackage{hyperref}
\usepackage{authblk}

\def\uu{\mbox{$u\overline{u}$}}
\def\dd{\mbox{$d\overline{d}$}}
\def\ss{\mbox{$s\overline{s}$}}
\def\cc{\mbox{$c\overline{c}$}}
\def\bb{\mbox{$b\overline{b}$}}
\def\ll{\mbox{$l^+l^-$}}
\def\qq{\mbox{$q\overline{q}$}}
\def\ud{\mbox{$u\overline{d}$}}
\def\du{\mbox{$d\overline{u}$}}
\def\cs{\mbox{$c\overline{s}$}}
\def\sc{\mbox{$s\overline{c}$}}

\def\pt{\mbox{p$_{\mathrm T}$}}
\def\ra{\mbox{$\rightarrow$}}
\def\gevc2{\mbox{GeV/c$^2$}}

\def\etainf{\mbox{$\eta_{\mbox{scale}}$}}

\def\plotsize{9cm}

\begin{document}

\title{Electroweak-QCD interference in  hadronic  vector bosons decays
  at the LHC}


\author{Cathy Ding \and Ben Fuller \and Eleanor Jones \and Aran Martin \and William Murray
}


\institute{C. Ding  \at Badminton School \and
  B. Fuller \at Sherborne School \and
  E. Jones \at University of Warwick \and
  A. Martin \at  The Royal Latin School \and
  W. Murray \at STFC Rutherford Appleton Laboratory \& University of Warwick
              \email{bill.murray@cern.ch}           
}

\date{Received: date / Accepted: date}

\maketitle

\begin{abstract}
The analysis of hadronic vector bosons decays at the LHC does not
normally consider interference with QCD \qq\ production. This paper explores the effect of this
interference on the reconstructed peak positions and rates  for several production modes. In particular, boosted vectors bosons and vector boson pairs are considered  for the first time. Shifts of several GeV/c$^2$ are seen in the positions of the $W$ and 
$Z$ boson's peaks, with a magnitude that depends strongly on
the kinematics involved. For boosted vector bosons in regions currently studied, the effects
are all found to be very small or negligible. If experiments were to
access lower transverse momenta, for example in the experimental
trigger systems, or in studies of low-recoil $VV$ decaying semileptonically, the effect of interference could be much larger.
\end{abstract}

\section{Introduction}

This paper presents a study of interference effects between electroweak production of  hadronic vector
bosons and QCD background at the LHC.
Such effects are neglected in the simulation strategy generally adopted by the LHC experimental collaborations, where vector bosons are simulated independently from QCD.  The goal of this paper is test the validity of that approach.
Early theoretical studies of interference~\cite{ranft79} explored its
impact  on total $W$ and $Z$ (collectively referred to here as $V$) cross-sections in $pp$ and $p\overline{p}$ collisions at energies below 1~TeV/c.
More details were explored theoretically at energies of 630~GeV/c and 1.8~TeV/c \cite{glover89} and these studies concluded that ``effects caused by the interference between electroweak and QCD amplitudes must be taken into account when data are compared with theoretical predictions''. A shift
in the masses of the resonance peaks of about 0.3~GeV/c$^2$ was noted and it was also observed that this effect was increased by experimental resolution.
First evidence for hadronic vector boson decay  was shown in 1990 by UA2 in two-jet decays \cite{Alitti:210698}, using 4.66pb$^{-1}$ of sp$\overline{\mathrm {p}}$s data. This result allowed for a 2.2~GeV mass shift in the experimental peak position from interference, based on Ref.~\cite{glover89}.
Pumplin \cite{PhysRevD.45.806} considered hadronic vector boson observation at the Tevatron as a mass calibration channel and remarked that interference causes a shift in mass of approximately 0.35~GeV/c$^2$ downwards. All of these studies were concerned with inclusive production at leading order in QCD.
No experimental evidence for this inclusive process has been published by  the LHC experiments.

There are several production modes of boosted hadronic vector bosons at the LHC. Baur \cite{baur:2006sn} considered hadronic $V+X$ production, finding it to be at the order of 1\% of the jet rate. Interest in hadronic vector bosons at the LHC was stimulated by the proposal to analyse boosted hadronic Higgs decays \cite{Butterworth:2008iy}. Inclusive $W \rightarrow q\bar{q}$ decays at high momentum have been used as a calibration in hadronic $X\rightarrow VV\ra \qq\qq$ searches \cite{Aad:2019fbh,Sirunyan:2019jbg}. Both CMS and ATLAS have also used boosted $Z\rightarrow \bb$ decays as a method of validating techniques to search for highly boosted Higgs bosons decaying to the same final state \cite{PhysRevLett.120.071802,ATLAS-CONF-2018-052}. Recently, the ATLAS collaboration performed a cross-section measurement in the $Z(\rightarrow b\bar{b})\gamma$ channel using 36~fb$^{-1}$ of data \cite{Aad:2019wdr} and this same channel has also been used to validate $b$-tagging developments \cite{Aad:2680245}. None of these studies considered interference.

Semileptonic $VV$ production provides experimentally accessible
signals. For example, the $VH, H\rightarrow \bb$ discovery papers by
ATLAS and CMS \cite{Aaboud:2018zhk,Sirunyan:2018kst} obtained clear
evidence for the $VZ, Z\rightarrow \bb$ process, summing over $V \ra
\nu\nu$, $l\nu$ and $\ll$. These searches used lower \pt\ thresholds
on the bosons than the inclusive ones discussed above, which it
will be shown gives rise to larger interference effects. 

The multiple distinctive features of top quark pairs allow the extraction of a clear hadronic $W$ boson signal, which has been used to validate jet
performance \cite{CMS-PAS-JME-16-003,ATLAS-CONF-2016-008} and  calibrate the top quark mass measurement \cite{Aaboud2019,Sirunyan2018}. A shift in the $W$ mass peak here could have a significant impact. This is considered briefly in the concluding section of this paper.

At this stage, it may be helpful to highlight an important point: if two amplitudes $a$ and $b$ contribute to a process, the cross-section is given by $\sigma = a a^* + b b^* + a b^* + ba^*$. If $a$ is the process under study and $b$ represents a background, then the fractional contribution of the interference term is at most $2 |b|/|a|$. Thus, (fractional) interference effects are expected to scale inversely with the square-root of the signal-to-background ratio.

Simulations of several processes are studied at leading order using Sherpa \cite{Gleisberg:2008ta} and the  procedure is discussed in section~\ref{sec:methodology}.
At higher order, there are additional
effects that are important in the prompt production of heavy quark
mesons at hadron colliders. This is especially evident in $\psi(2s)$
production, where at high \pt\ even Next-to-Leading-Order QCD (NLO)
calculations under-predict the 
rate by a factor of two \cite{Cacciari:2012ny}.  
The OpenLoops
\cite{Cascioli:2011va} package was explored as an NLO generator, but
no relevant states are currently interfaced to Sherpa. 
The results for the inclusive   vector bosons, for comparison with previous literature, are given in section~\ref{sec:inclusive}, and, in addition, the analysis methodology is introduced there. This is then applied to inclusive boosted $W$ and $Z$ bosons in section~\ref{sec:boosted}, $V+\gamma$ in section~\ref{sec:vgamma} and $VV$ in section~\ref{sec:vv}. Section~\ref{sec:discussion} contains a discussion of the results and prospects for experimental verification.

\section{Methodology}
\label{sec:methodology}

The results presented in this paper are obtained using the Sherpa
2.2.8 package with Photos \cite{Schonherr:2008av} and normally the
NNPDF3.0 \cite{Ball:2014uwa} Parton Distribution Function (PDF) set, with MMHT2014nlo68cl
\cite{Harland-Lang2015} used for certain studies. The Comix
\cite{Gleisberg:2008fv} matrix element generator is used throughout to
generate leading order cross-sections. As such, $s$-channel
gluon-mediated processes are pure colour octets and do not interfere
with vector bosons. Furthermore,
the parton shower, hadronization and underlying event are mostly
disabled to save computing time, so the resulting output is a small
set of partons only. For a small number of examples, the parton shower
is enabled and the DelphesMC \cite{deFavereau2014} simulation is
employed, using the `ATLAS' detector card to approximate the influence
of a detector on the results. These cases require a jet treatment,
discussed in  section \ref{sec:systematics}.

Sherpa uses the concepts of a `process' and a `selector'. The process
defines the initial and final state particles, with additional options
such as defining intermediate states or the order of the electroweak
coupling at the amplitude level. The aim of this paper is to study the
resonant vector boson peaks, and their interference with the continuum
background, but there is no simple way to include all non-resonant
electroweak processes in the background. Instead, the three samples
generated are a total, a background (which is referred to as QCD
throughout), with the minimum electroweak order to reach the specified
final state, and the electroweak signal, where two additional orders
in the electroweak coupling are required corresponding to the creation
and decay of an internal electroweak boson. So, for example, $\gamma
\ud$ is generated at order one and order three in the electromagnetic
coupling, though most states involve order zero and order two. The
selector imposes kinematic selections on the outgoing 
particles. In this paper, the selectors applied are:
\begin{itemize}
\item a mass window on the relevant quark-antiquark pair of
  50~GeV$/c^2 < m < 130$~GeV$/c^2$. This wide window enables studies
  involving the experimental resolution.
\item all final state particles must have a \pt\ greater than
  25~GeV/c. This mimics experimental selection and stabilises the
  Sherpa integration.
\item all charged objects must have $|\eta| < 2.5$. This mimics
  central detector acceptance and also serves to stabilise the Sherpa
  integration in some cases.
\item when studying boosted bosons, the quark-antiquark pair is
  required to pass a \pt\ selection. In the case where there are three
  coloured objects in the final state, this requirement is instead
  placed on the leading one, which, at this matrix element level, has
  identical \pt\ to the recoiling pair.
\end{itemize}

Simulation is done at 13~TeV centre-of-mass energy for proton-proton
collisions. A full `Run.dat' file defining one process can be seen in
appendix~\ref{sec:run}. Within this, it can be seen that the output is
in hepMC format. DelphesMC is then used to reformat this into a file
that can be processed by root 6.16 \cite{Brun:1997pa}.

The analysis extracts the interference component as the QCD background
and electroweak signal subtracted from the total and comparisons are
made between the pure electroweak signal and the electroweak plus the
interference term ($EW+I$). The invariant mass of the \qq\ pair is
examined. For processes where there are only two quarks in the final
state, there is only one combination to form the invariant
mass. However, for processes with a \qq\ pair plus an extra quark in
the final state, there are two possible combinations and  
both are considered.

 \section{Vector boson production at rest}
   \label{sec:inclusive}

In this section, the simplest case of $W$ and $Z$ production with no
recoil, $p\overline{p} \rightarrow V$, is considered. This mode shows
clear interference effects and so is also used to check the
impact of the selection, the experimental resolution and additional
gluon radiation.

\begin{figure}[htb] 
 \centering \includegraphics[width=3.8cm]{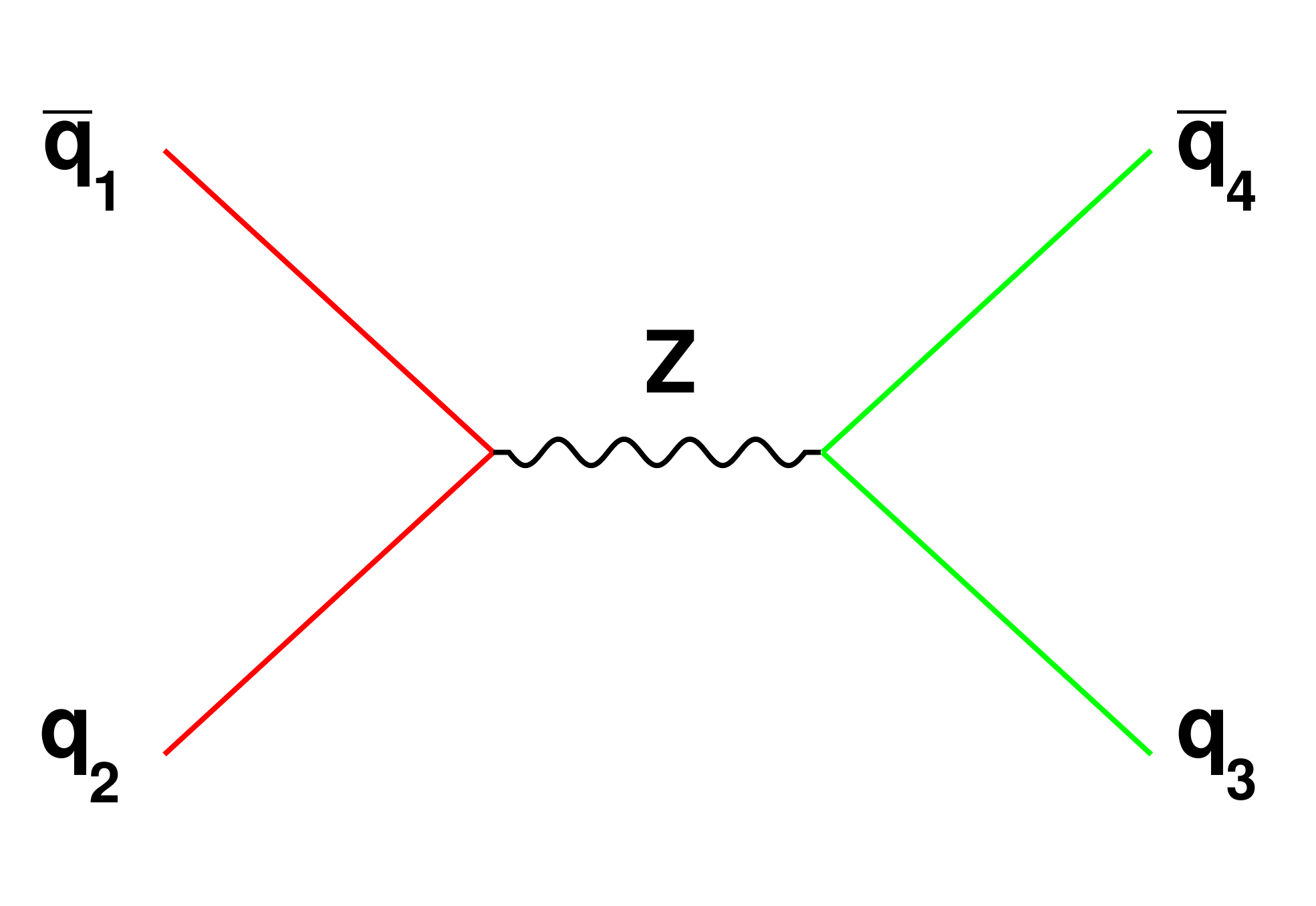}
 \includegraphics[width=3.8cm]{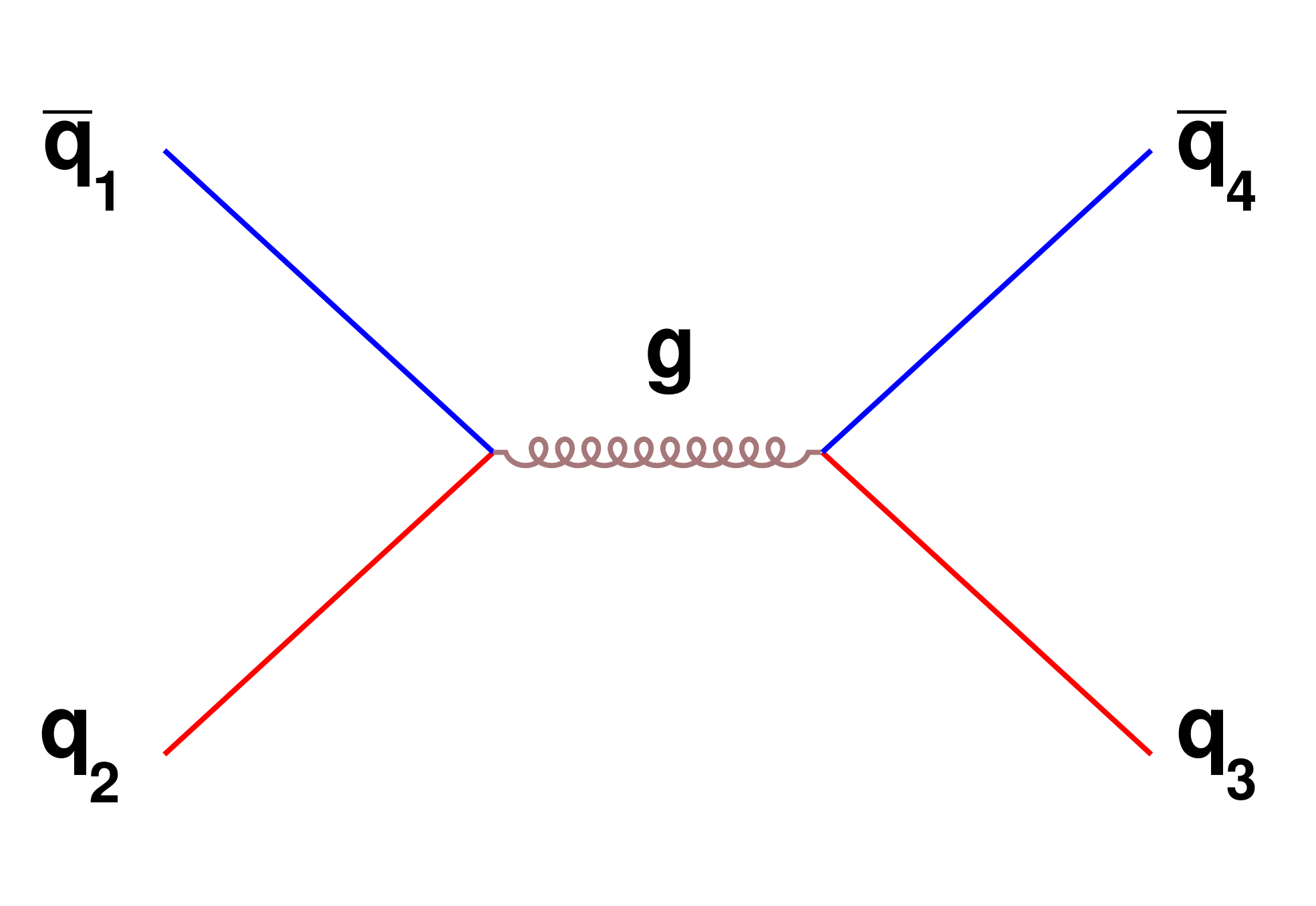}
 \includegraphics[width=3.8cm]{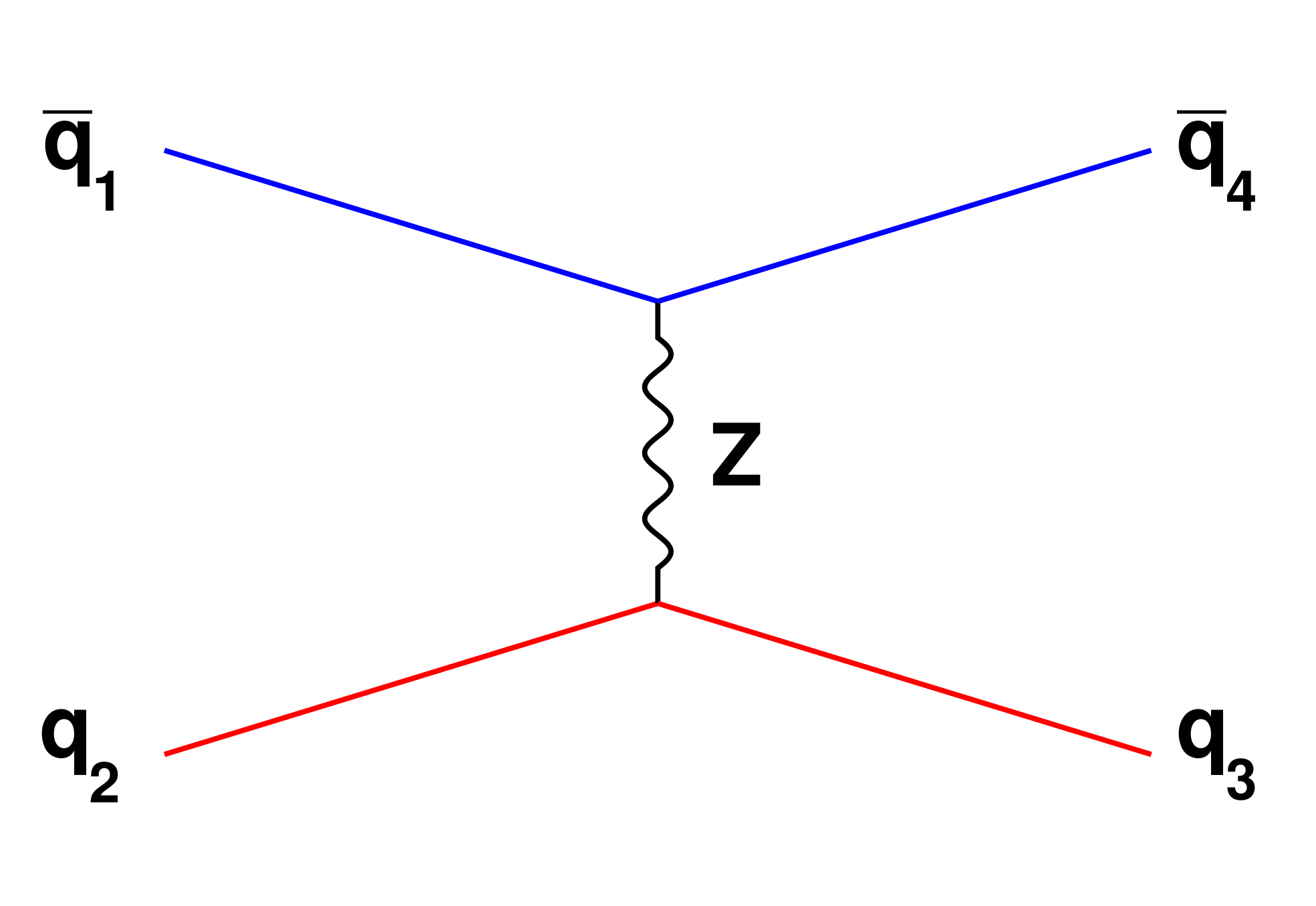}
 \includegraphics[width=3.8cm]{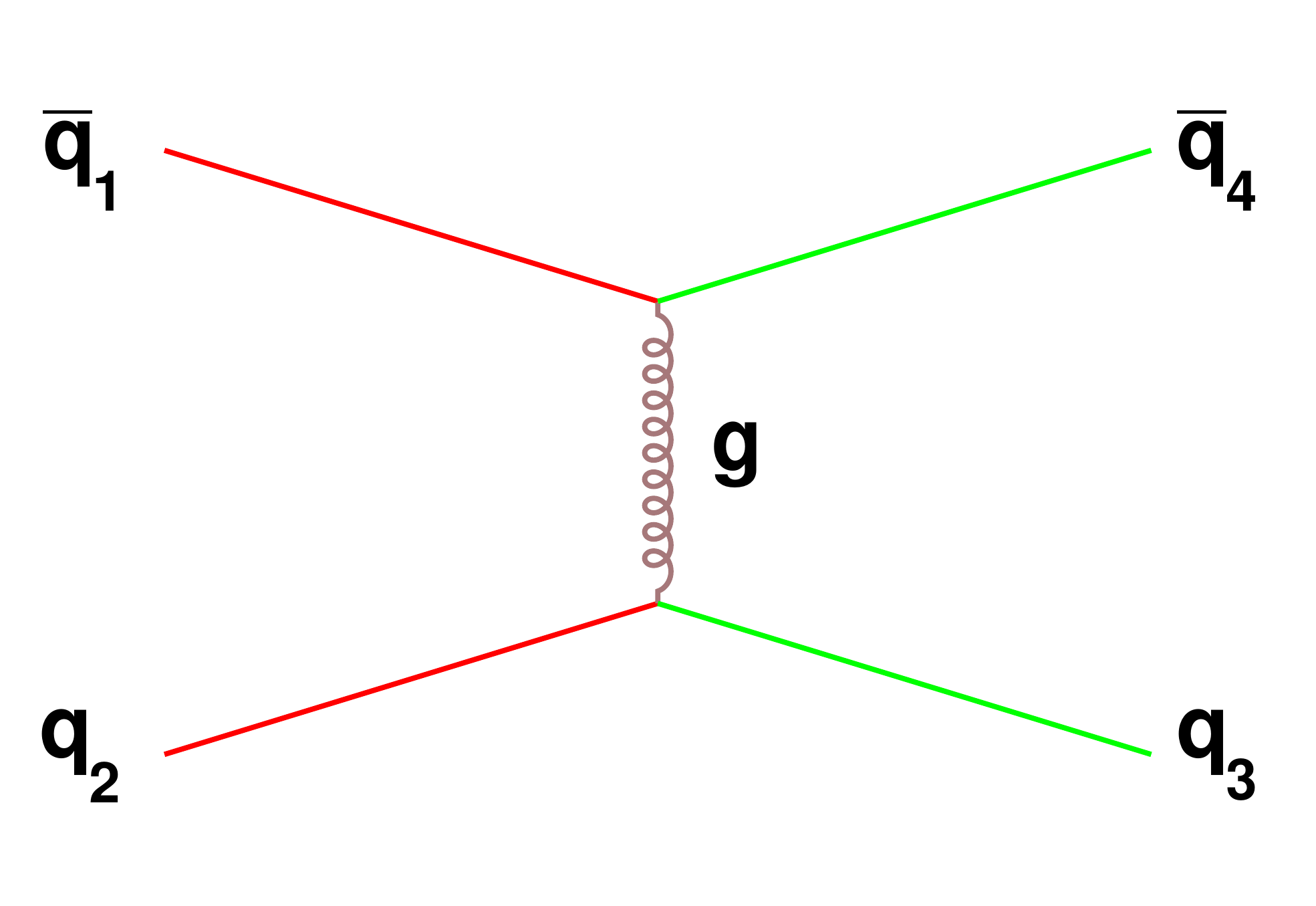}
 \vskip -0.2cm
 \caption[]{The leading order \qq $\rightarrow$ \qq\ diagrams, with
   $s$- and $t$-channel $Z$ and gluon processes. The $Z$ could be
   replaced by a $W$ or photon.}
 \label{fig:qqqq} 
 \end{figure}

At leading order, $Z$ or $W$ boson production can only proceed through
quark-antiquark annihilation. The experimental background has large
contributions from processes involving gluons in the initial or final
state. However, interference only occurs between processes with
identical initial and final states, so these background processes with
gluons will not contribute. To study interference, it is sufficient to
focus on $\qq \ra \qq$. The total jet-jet cross-section is two orders
of magnitude larger than the \uu\ cross-section so this represents a
considerable calculation simplification. Figure~\ref{fig:qqqq} shows
the Feynman diagrams for the electroweak and QCD processes.
Only the $s$-channel electroweak process produces a resonant mass
signature in the \qq\ final state and it is a colour-singlet. However,
the similar $s$-channel QCD diagram is a colour octet and, as a
result, it does not interfere. This does not hold in general for the
$t$-channel gluon exchange process. Therefore, interference requires
identical initial and final state particles. However, while the
experimentalist has some possibility to distinguish final state
quarks, all \qq\ initial states must be included. This means, for
example, that the $t$-channel QCD diagrams for a \bb\ final state are
suppressed compared to the \uu\ final state as a result of the
relatively small $b$ quark PDF. Consequently, it is to be expected
that the apparent interference is reduced for heavy quarks.

\subsection{Inclusive $Z$ production}
\begin{figure}[htb] 
\centering \includegraphics[width=\plotsize]{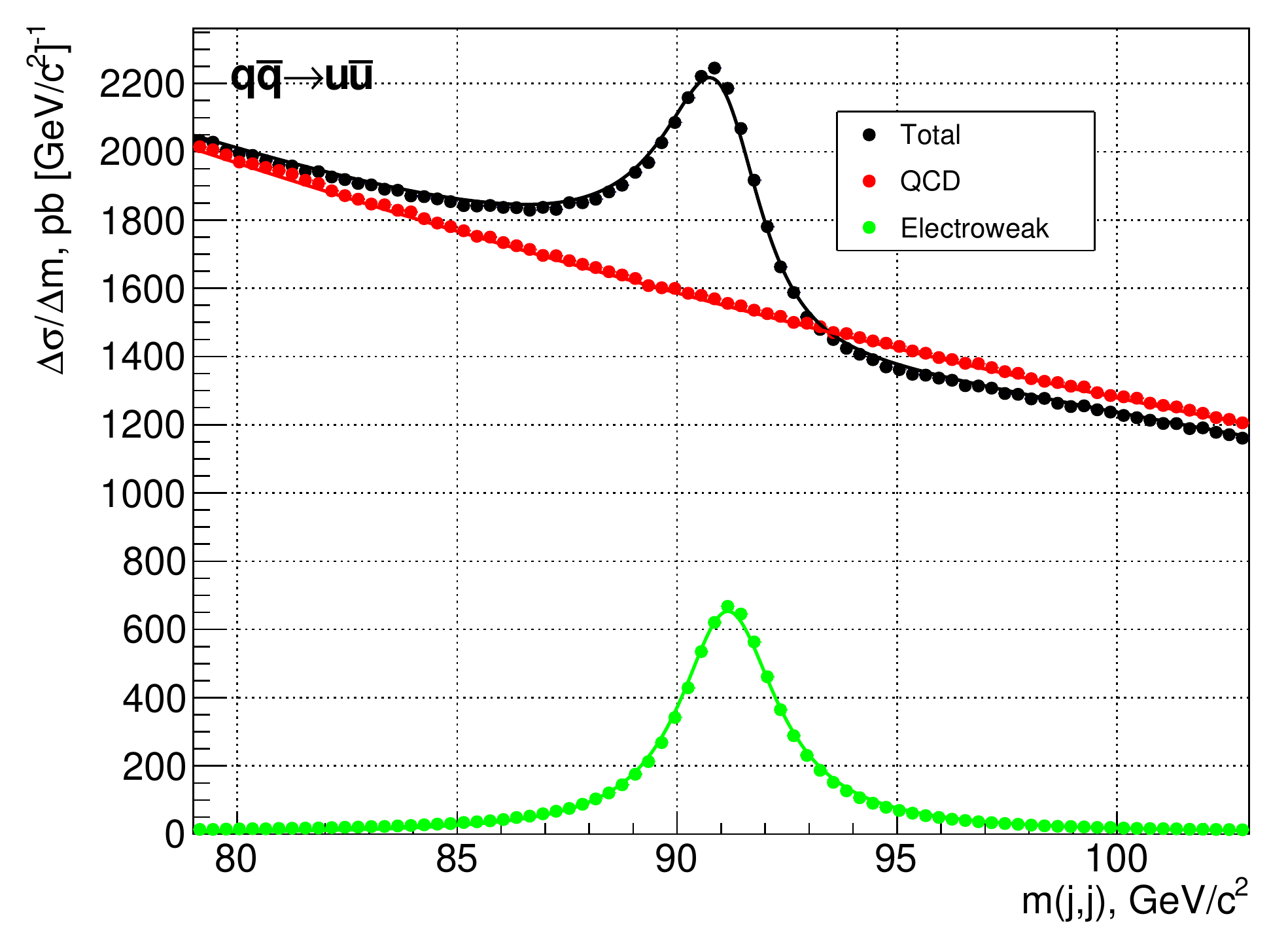}
\includegraphics[width=\plotsize]{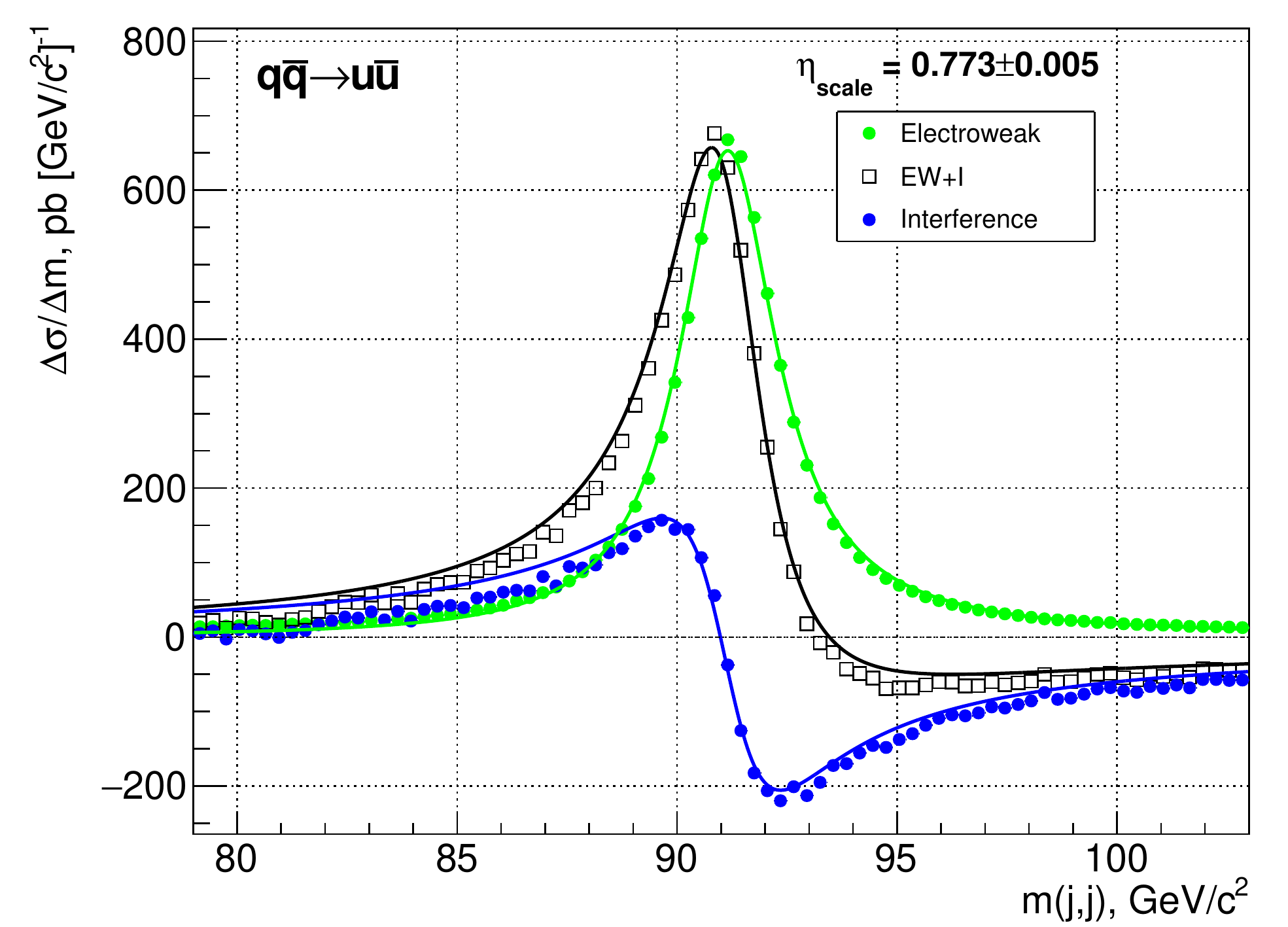}
\vskip -0.2cm
\caption[]{The process $q \overline{q} \rightarrow \uu$. Left shows
  the pure electroweak (pale/green), pure QCD (dark/red) and total
  (black) cross-sections. Right has the same electroweak component,
  but the hollow black points show the total minus the QCD component,
  and the dark (blue) is the interference term. The fit is discussed
  in the text.}
\label{fig:z-rest} 
\end{figure}

To study $Z$ production, the five final state quark-pairs are
simulated separately. The differential cross-section for $\qq\ra\uu$ is shown
in figure~\ref{fig:z-rest}, with the QCD and electroweak channels
compared with the combined process.
There are sizeable interference effects in the mass spectrum, with an
enhancement below the peak and destructive interference above. These
effects are largest about $\pm \Gamma/2$ from the pole mass, but they
also induce a small shift in the peak position. They are modelled
using the sum of a relativistic Breit-Wigner function and an arbitrary
constant amplitude:
\begin{equation}
\mathcal{A} = \frac{2 m_0 \sqrt{\kappa (m/m_0)^3}}{ (m^2-m_0^2) + i
  m_0 \Gamma} +\sqrt{b} +\sqrt{c}i
\end{equation}
In this formula, $\kappa$ is the normalisation of the Breit-Wigner, $m$ is the
invariant mass of the system, $m_0$ and $\Gamma$ are the particle mass
and width, respectively. $b$ and $c$ are the coefficients of the real
and imaginary amplitude components of an interfering background,
respectively.
This formula neglects variation in the interfering amplitude with
mass. More importantly, in the identification of the $b$ and $c$ terms
with the interfering part of the QCD contribution, it ignores
$t$-channel electroweak processes. The motivation for the $m^3$
dependence of the Breit-Wigner rate is heuristic.

The cross-section is not just $\mathcal{A}\mathcal{A}^*$, but has an
additional term, $d(m)$, which is a non-interfering background component  modelled with a second-order polynomial dependence on $m$:

 \begin{eqnarray}
\frac {\partial \sigma}{\partial m} & = & \frac{2m_0 \sqrt{\kappa
    (m/m_0)^3}}{ (m^2-m_0^2)^2 + m_0^2 \Gamma^2} \left( 2m_0
\sqrt{\kappa(m/m_0)^3} + \sqrt{b} (m^2-m_0^2) + \sqrt{c} m_0 \Gamma \right)~  + b
+ c + d(m)~ \mathrm {pb}~[\gevc2]^{-1}.
  \label{eqn2}
\end{eqnarray}

These nine parameters are simultaneously fit to the three
distributions in figure~\ref{fig:z-rest} (left), with terms dropped
for the partial distributions where they do not contribute. The
extracted values are $m_Z = 91.125\pm0.001$~GeV/c$^2$ (c/f
91.19~GeV/c$^2$ used in simulation) and $\Gamma_z =
2.687\pm0.003$~GeV/c$^2$ (c/f 2.50~GeV/c$^2$ used in simulation). The
increase in $\Gamma_z$ with respect to the simulation is due to the
fit attempting to describe the $t$-channel exchange in the electroweak
component.

The fit also finds
$b=202\pm2~\mathrm{pb}~[\gevc2]^{-1}$
and
$c=2.2\pm0.2~\mathrm{pb}~[\gevc2]^{-1}$;
a sizeable
real term and a significant imaginary one, which is not expected and
is interpreted as being due to the $t$-channel electroweak exchange. The
size of the interfering QCD component $b$ is 
approximately one eighth of the total QCD contribution to \qq\ra\uu\ at $m_Z$. The
components of the fit are shown in figure~\ref{fig:z-rest} (right) and
it can be seen that the approximate features are captured, but the
limitations  are clearly visible.
The fit is informative, but due to the missing
electroweak $t$-channel, it is not quantitatively reliable and simpler
methods are generally used for the rest of this paper.

A parameter, \etainf, characterising the scaling of the signal strength
is defined as the integral of the $EW+I$ component in a window of
$\pm$10~GeV/c$^2$ around the nominal mass, divided by the same
integral for the electroweak term alone. This is the same as the definition of
$\eta$ presented by the `New Physics' Working Group
\cite{Brooijmans:2018xbu}, but the integration limits have been
reduced from the $\pm 10\Gamma$ window to one of $\pm$10~GeV/c$^2$ which is 
comparable to the typical experimental resolution. This revised
definition has less sensitivity to the $t$-channel component. The
other parameter extracted is the shift of the peak position. This is
established by fitting a Gaussian to the observed data, in a mass
range defined by the half-maxima of the distribution. The shifts in
the peak positions due to interference are evaluated for all the $Z$
decay modes and given in table~\ref{ta:species}.
\begin{table}[htp]
\begin{center}
\begin{tabular}{l|ccccc} & \dd & \uu & \ss & \cc & \bb \\
 \hline signal-to-background & 0.60 & 0.43 & 1.08 & 1.18 & 2.6 \\ Shift, GeV/c$^2$ &
 $-0.384\pm$0.014 & $-0.405\pm$0.009 & $-0.202\pm$0.010 &
 $-0.135\pm$0.011 & $-0.066\pm$0.007 \\
 \etainf & 0.86$\pm$0.01 &
 0.77$\pm$0.01 & 0.95$\pm$0.01 & 0.91$\pm$0.01 & 1.009$\pm$0.003 \\
\end{tabular}
\caption{Interference effects on the $\qq\ra W$ peak by final state quark flavour. Signal-to-background is defined at the resonance peak without experimental resolution and using only the initial and final states quoted, with negligible statistical error.
\label{ta:species}}
\end{center}
\end{table}
The peak shift, averaged over the final state quark species, is
-0.23$\pm$0.01~GeV/c$^2$, similar to the $0.3 - 0.35$~GeV/c$^2$ shifts
reported in the previous work discussed in the introduction, but
smaller with the higher beam energy, as expected. The magnitude of the
effects falls as the quark mass rises, since the PDF density reduces,
and so the non-interfering $s$-channel QCD process becomes more
dominant. 
 
 These shifts are for the theoretical peak in the cross-section. Estimating
 the expected change in the measured peak position requires allowing
 for detector resolution. This was discussed by the `New Physics'
 Working Group \cite{PhysRevD.86.073016}, and Baur and Glover \cite{glover89},
 where the latter adopted a 10\% Gaussian smearing to approximate
 detector resolution effects. The same 10\% is used here, noting that
 ATLAS found a mass resolution between 8\% and 16\% for boosted dijets
 with \pt\ between 300 and 1000~GeV/c \cite{Aad:2019fbh} and CMS
 shows between 7.5\% and 10\% for $W$ and $Z$ from the decay of 1200 -
 5000~\gevc2 bosons \cite{Sirunyan:2019jbg}.
 
\begin{figure}[htb] 
\centering
\includegraphics[width=\plotsize]{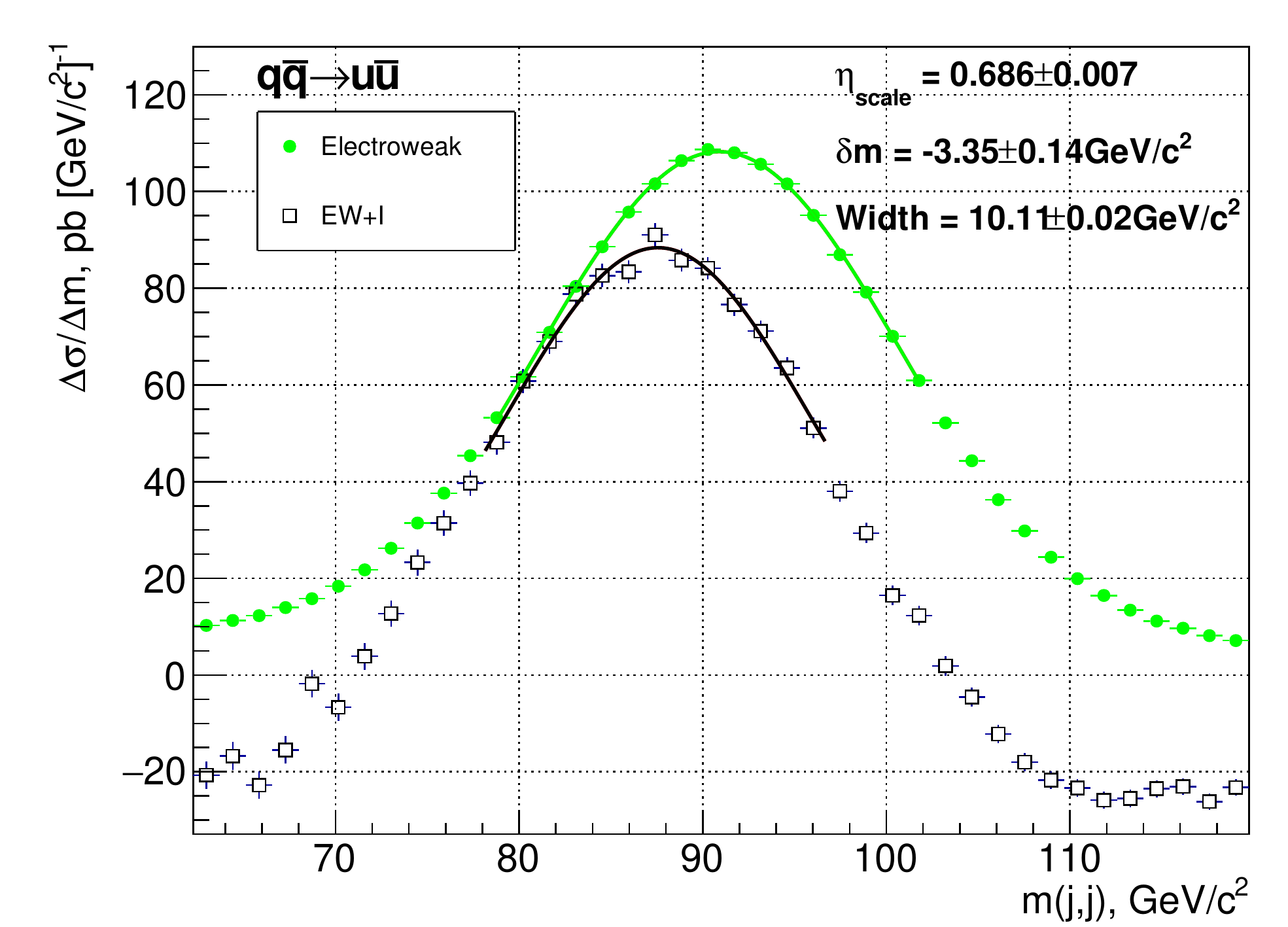}
\includegraphics[width=\plotsize]{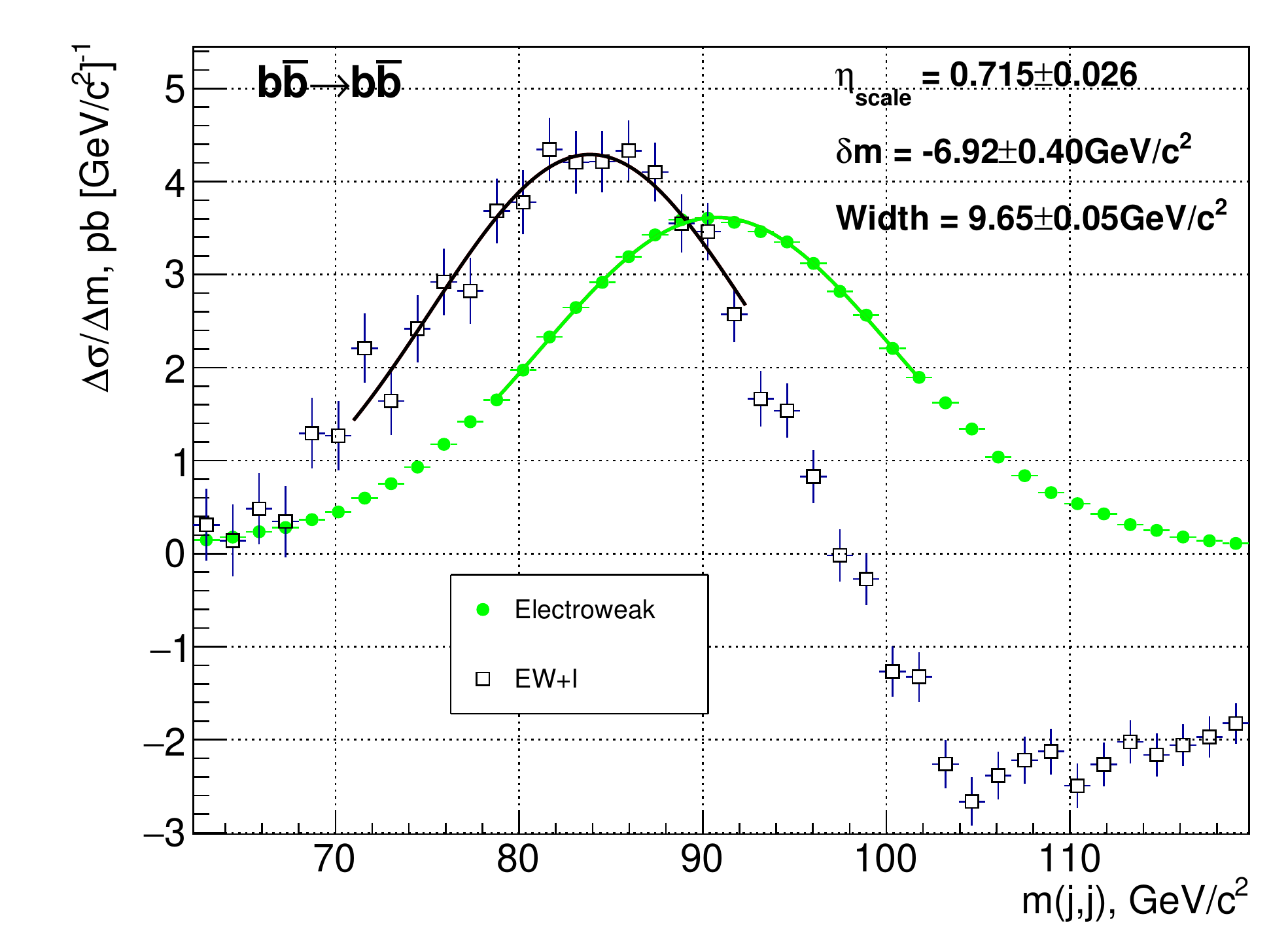}
\vskip -0.2cm
\caption[]{Left: The process $q \overline{q} \rightarrow
  u\overline{u}$ after smearing by an assumed 10\% resolution. The
  electroweak component is  pale/green, but the hollow black
  points show the $EW+I$. Right is the same distribution for $b
  \overline{b} \rightarrow   b\overline{b}$, showing a significantly
  larger effect.}
\label{fig:z-rest-det} 
\end{figure}
Figure~\ref{fig:z-rest-det} (left) shows that after smearing,
the $EW+I$ component is negative both above and below the peak, unlike
what is seen in the original distribution, as the smearing makes the
$t$-channel contribution fractionally larger. The $t$-channel imposes a
 decrease in differential cross-section across the mass spectrum, and effectively a baseline shift under the peaks.
This  gives rise to the reported reduction in \etainf, 0.69. If the background
had been extracted from a fit to sideband  data,  this reduction in
cross-section  would
not be seen.
The fit to the electroweak alone yields good
agreement with the $Z$ mass, while the $EW+I$ distribution is
down-shifted by $3.35\pm0.14$~GeV/c$^2$, as a result of the
interference.

This can be compared with the shift in the $\qq \ra Z \ra \uu$ peak position
before smearing of  $-0.405\pm0.009$~GeV/c$^2$ which shows the
experimental resolution increases the shift by an order of
magnitude. The larger number can be compared with the systematic error
obtained by the experiments on jet energy scale, which in the best
measured regions are around 1\%
\cite{Aaboud:2017jcu,Khachatryan:2016kdb} or even a little less. It
can be seen that the use of vector boson resonances as a standard
candle for mass scale is potentially problematic.

The analysis is repeated for only the subprocess $\bb \ra Z \ra \bb$ where large effects are expected as the initial and final state quarks are identical. 
This finds a  peak shift of $-0.81\pm0.02$~\gevc2,  larger than any shift reported  in table~\ref{ta:species}.
This demonstrates that the small interference in the \bb\ final state is because its initial state which is dominated by the  non-interfering lighter-quarks. Figure~\ref{fig:z-rest-det} (right)  shows the mass distribution  
for $\bb \ra \bb$ after
smearing by 10\%, where  the change in the peak position
is $-6.9\pm0.4$ \gevc2. Measuring this distribution would require
knowledge of  the initial state quarks, which is not experimentally accessible.

 \subsubsection{Variation of procedure}
 \label{sec:systematics}

This section discusses the stability of the results presented
above as the physics assumptions involved are varied.

 The sensitivity of the peak position to the PDF uncertainties has been
 explored using the 50 PDF eigenvector variations of
 MMHT2014nlo68cl. \footnote{Attempting to use the PDF variations of
   NNPDF30NNLO caused Sherpa termination for memory overflow on a 128
   GB RAM computer.}  The quadratic sum of the resulting shifts in the
 $\qq\ra Z \ra \uu$  peak position is 0.006~\gevc2, corresponding to a 1.5\% uncertainty on
 the shift. PDF uncertainties are therefore considered negligible.

Variation of the \pt\ threshold used to select quarks in the process
generation has also been studied. This was normally set to 25~GeV/c,
but selections of  20 and 30~GeV/c have been investigated. The change in the
pure electroweak cross-section is $\pm$10\%, but for the QCD it is a
factor of two at 90~GeV/c$^2$. The effect is much larger for QCD
because the $t$-channel production tends to have large angular
separation. Thus, lower \pt\ thresholds enlarge the QCD sample in the kinematic region where
the colour structure can interfere. The result is that the peak shift
has quite a strong dependence on the \pt\ selection, as seen in
table~\ref{ta:pt_eta}.

   \begin{table}[htp]
   \begin{center}
     \begin{tabular}{lcccccc}
       Quark \pt & \multicolumn{2}{c}{Inclusive } &
       \multicolumn{2}{c}{$ | \Delta \eta| < 1$} &
       \multicolumn{2}{c}{$ | \Delta \eta| > 2$} \\ threshold &
       cross-section & peak shift& cross-section & peak shift&
       cross-section & peak shift \\ \hline 20 GeV/c & 1882 pb &
       $-0.465\pm0.017$\gevc2 & 954 pb & $-0.280\pm0.017$\gevc2 & 283
       pb & $-0.933\pm0.045$\gevc2 \\ 25 GeV/c & 1741 pb &
       $-0.405\pm0.009$\gevc2 & 945 pb &$-0.309\pm0.008$\gevc2 & 156
       pb & $-0.737\pm0.056$\gevc2 \\ 30 GeV/c & 1564 pb &
       $-0.351\pm0.009$\gevc2 & 945 pb & $-0.302\pm0.010$\gevc2 & 0.2
       pb & n.a. \\ \hline
\end{tabular}
 \caption{The  electroweak cross-section accepted  within a window of
   89 to 93 GeV/c$^2$ in $m_{\uu}$, selecting the $Z$ peak, as well
   as the fitted shift in that peak due to interference. These are
   given for  various
   quark \pt\ thresholds and diquark $\Delta\eta$ selections. A
   \pt\ threshold of 30 GeV/c combined with a $|\Delta\eta| > 2$
   essentially excludes the $Z$ boson peak, hence there is no entry
   for that region.
   \label{ta:pt_eta}
   }
 \end{center}
 \end{table}
The dependence of the mass shift on the $|\Delta \eta |$ between the
quarks is also explored in table~\ref{ta:pt_eta}. It can be seen that
for events where the quarks have similar $\eta$, the \pt\ threshold has
little impact, while at large  $|\Delta\eta|$, where the  \pt\ can be lower, 
 the threshold is much more important.
Thus,  the  results have a significant dependence on the
kinematic selection used.

\begin{figure}[htb] 
 \centering
 \includegraphics[width=\plotsize]{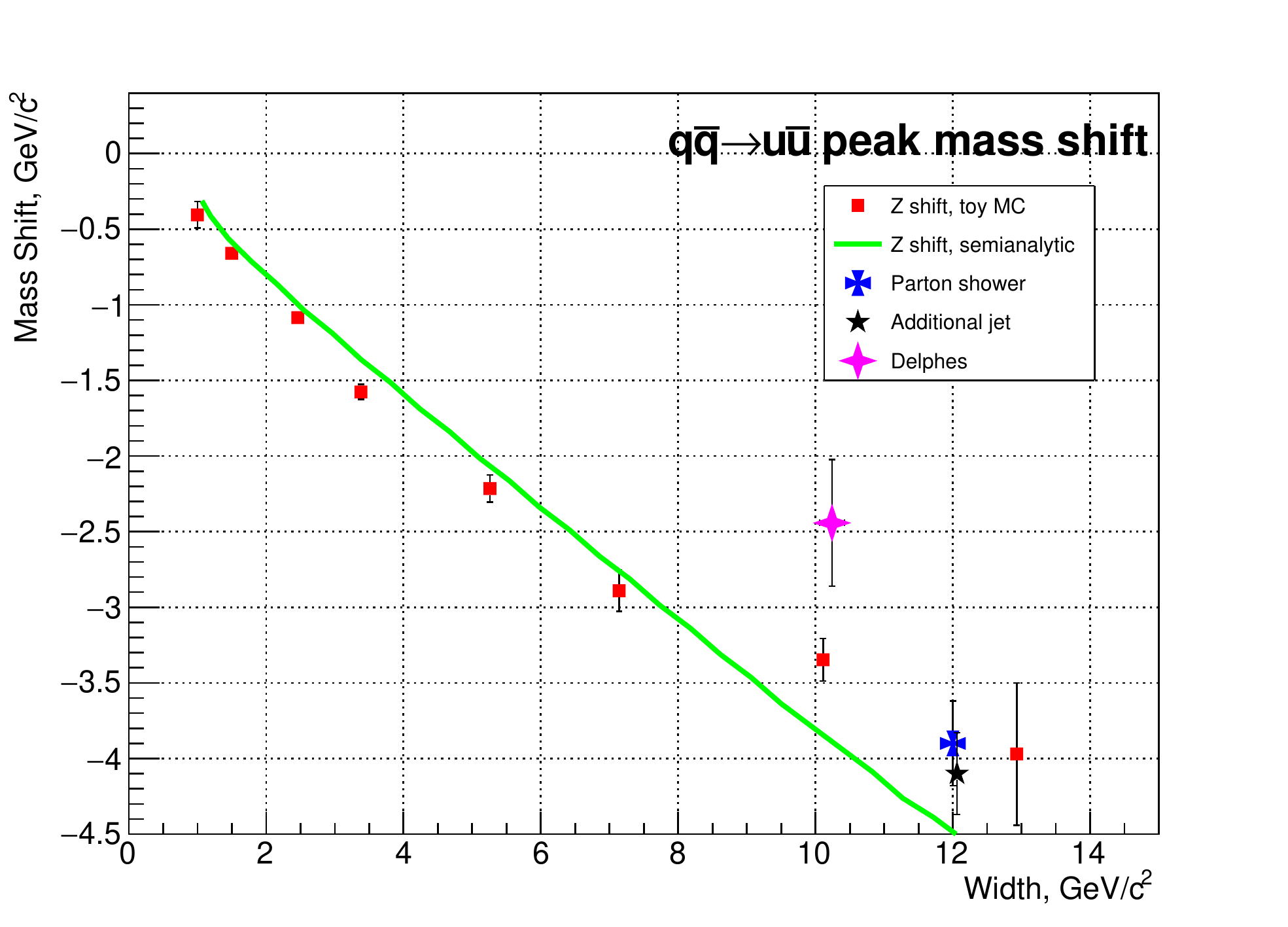}
 \includegraphics[width=\plotsize]{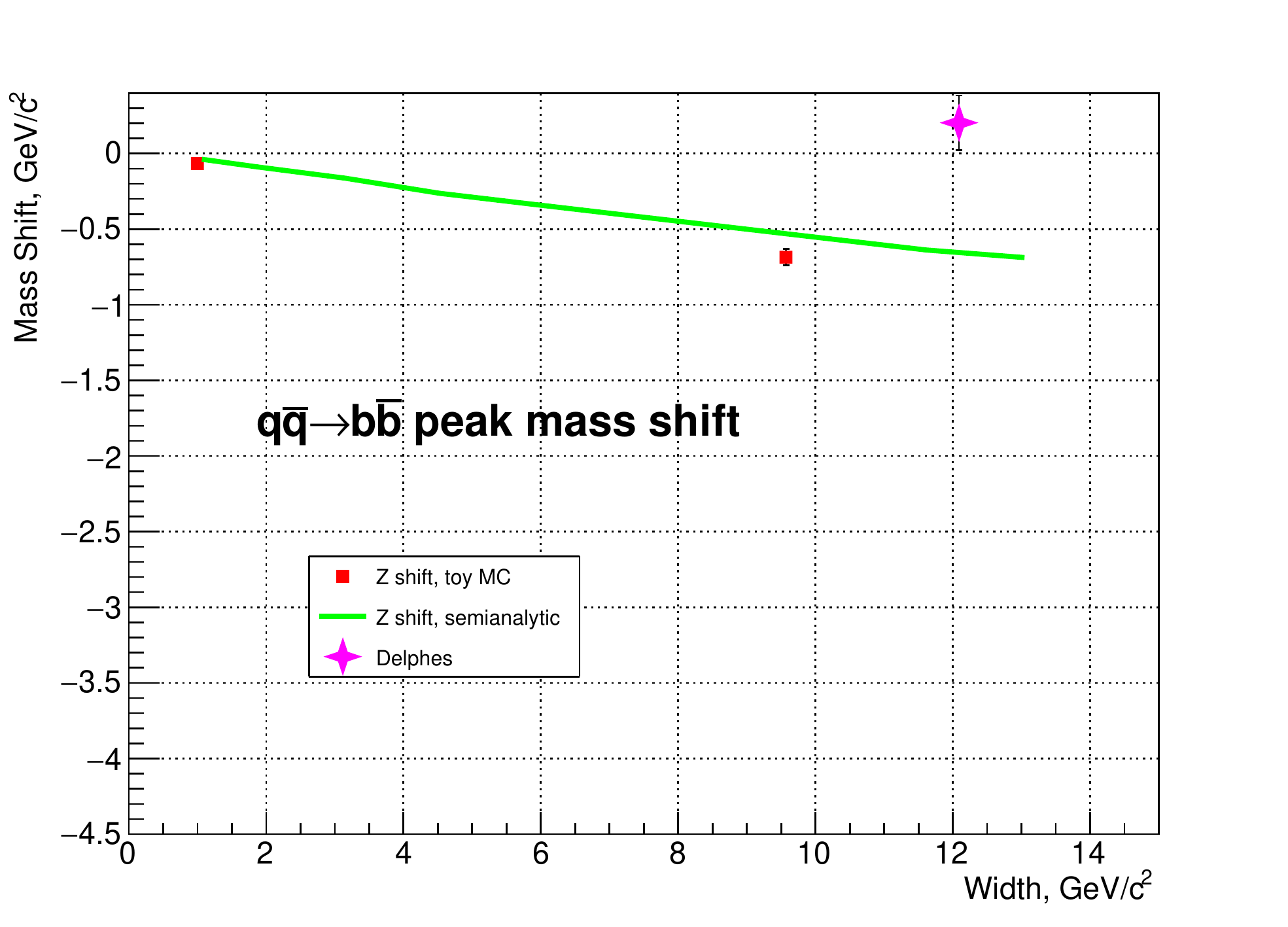}
 \vskip -0.2cm
 \caption[]{The shift in the peak positions of Gaussian distributions fitted to
   \uu\ (left)and \bb\ (right) peaks, plotted against the fitted width.
   The squares show the impact of different smearing
   resolutions, and the star symbols the effect of gluon radiation or
   Delphes simulation as discussed in the text. The continuous lines
   show the impact of a numerical smearing applied to the fitted
   interference formula.}
 \label{fig:shifts} 
 \end{figure}

 The shift is also a function of the assumed resolution and a smearing
 width varying from 0\% to 13\% is tested in figure~\ref{fig:shifts}. This is
 compared with the shift extracted from convolving equation~\ref{eqn2}
 with a Gaussian of variable width. The match is good, apart from a
 small offset, until the width reaches 10\%, where the impact of the
 $t$-channel interference, not included in the equation, becomes
 important.

The figure also shows the result of three progressively more complete
ways of introducing resolution: allowing a parton shower, allowing a
parton shower and an additional hard parton in the matrix element, and
a full hadronization plus the use of Delphes detector simulation, with
the ATLAS detector simulation employed. 
The resulting particles are all analysed
using Delphes, with the anti-k$_{t}$ \cite{Cacciari:2008gp} jet
algorithm employed to order the resulting particles. The events are
required to have precisely two jets above the 25~GeV/c \pt\ threshold
and, in the case of the Delphes simulation, the generator threshold is
lowered to 15~GeV/c, with the reconstructed jets still required to exceed 25~\gevc2,  to allow for threshold effects.  The two jets are
required to be coplanar to 0.1~radians, which reduces events where a
lot of energy is missed from the jet.  Finally, in the two cases where
Delphes was not used, a 10\% detector smearing is applied to the mass.

The resulting peak widths are 12.0~\gevc2 for the parton shower,
12.1~\gevc2 with additional hard radiation and 10.2$\pm$0.2~\gevc2 for
the Delphes simulation. This happens to match almost exactly the width
resulting from the 10\% smearing applied to the quarks.  The shift in
the peak position is 3.91$\pm$0.28~\gevc2 for the parton shower,
4.10$\pm$0.27~\gevc2 for the parton shower plus possible hard jet, and
2.42$\pm$0.42~\gevc2 for the Delphes events. These numbers can be
visualised in figure~\ref{fig:shifts} (left), which shows a generally
consistent pattern, with the Delphes results somewhat higher.  This
has been traced to reduced acceptance for events from Delphes
simulation at large $|\Delta \eta|$, presumably because such events
have a \pt\ close to the acceptance threshold and can easily fall
below it as a result.  This is the region where the interference
effects are larger, see table~\ref{ta:pt_eta}.  \footnote{These Delphes studies
in particular consumed many thousands of CPU hours and over a hundred
terabytes of disk.}

The right hand side of figure~\ref{fig:shifts} shows a comparison of
no smearing, 10\% smearing and parton shower plus Delphes for the
$Z\ra\bb$ case. A very similar pattern to the $\uu$  is seen, with the
effects scaled down by a factor of six.

\subsection{Inclusive $W$ production}

Inclusive $W$ production is analysed in a manner very similar to the $Z$, but because the  $W$ boson is charged  it does not have any  $s$-channel QCD background. However, this background did 
not create interference in the $Z$ case and so the actual effects
observed are similar.  Sherpa is run in its default form, with a
diagonal CKM matrix, so the possible signatures are \ud, \du, \cs\ and
\sc. The peak shifts and scale changes are given in
table~\ref{ta:wspecies}.

 \begin{table}[htp]
 \begin{center}
 \begin{tabular}{l|cccc}
   & \ud & \du & \cs & \sc \\
 \hline signal-to-background & 1.97 & 2.05 & 5.2 & 4.6 \\
    Shift, GeV/c$^2$ & -0.287$\pm$0.005
    & -0.242$\pm$0.006 & -0096$\pm$0.005 & -0.099$\pm$0.005
    \\ \etainf &
   1.018$\pm$0.003 & 1.027$\pm$0.004 & 0.995$\pm$0.002 &
   1.039$\pm$0.003\\
\end{tabular}
\caption{Interference effects on the $\qq\ra W$ peak by final state quark flavour. Signal-to-background is defined at the resonance peak without experimental resolution and using only the initial and final states quoted, with negligible statistical error.
 \label{ta:wspecies}}
 \end{center}
 \end{table}
 
 The shifts are found to be similar to, but smaller than, those in the $Z$ case, and
 the reduction of the effects for second-generation quarks is clear,
 as was also the case for the $Z$ boson.

   \section{Boosted vector bosons}
   \label{sec:boosted}

This section describes the dominant boosted production modes, $Vq$ and $Vg$. These modes have
different interference properties, but experimentally, they are almost
impossible to distinguish. Recoil against photons or other vector bosons is
considered later.

Reference~\cite{Aad:2019fbh} used a
\pt\ selection of 600~GeV/c, but
reference~\cite{PhysRevLett.120.071802} clearly shows that an
inclusive hadronic $W$ 
peak could be measured with a \pt\ selection of 450~GeV/c.
ATLAS and CMS used unprescaled jet trigger thresholds of approximately 400~GeV/c in run 2 at the LHC. In order to imitate the experiments, a \pt\ threshold of 400~GeV/c is used as the default here.
The extraction of a hadronic $Z$ boson signal experimentally often relies on $b$-quark identification to reduce QCD backgrounds suppress the larger $W$  peak, so this is given particular attention here.
 
\subsection{Inclusive Boosted $W$}

The processes requested for $W\ra\ud$ are $\qq \ra \ud g$, 
$q g \ra \ud q$ or 
$\overline{q} g \ra \ud \overline{q}$. 
It is required that there is a
\ud\ quark pair with a mass of 50 to 150~GeV/c$^2$, and as none of the
diagrams allow for two identical  quarks this is unambiguous.
In addition, at least one of the partons must have a \pt\ above the 400~GeV/c threshold. 
In general, this means that the highest \pt\ parton is recoiling
against the \ud\ pair. 

 \begin{figure}[htb] 
 \centering 
 \includegraphics[width=3.8cm]{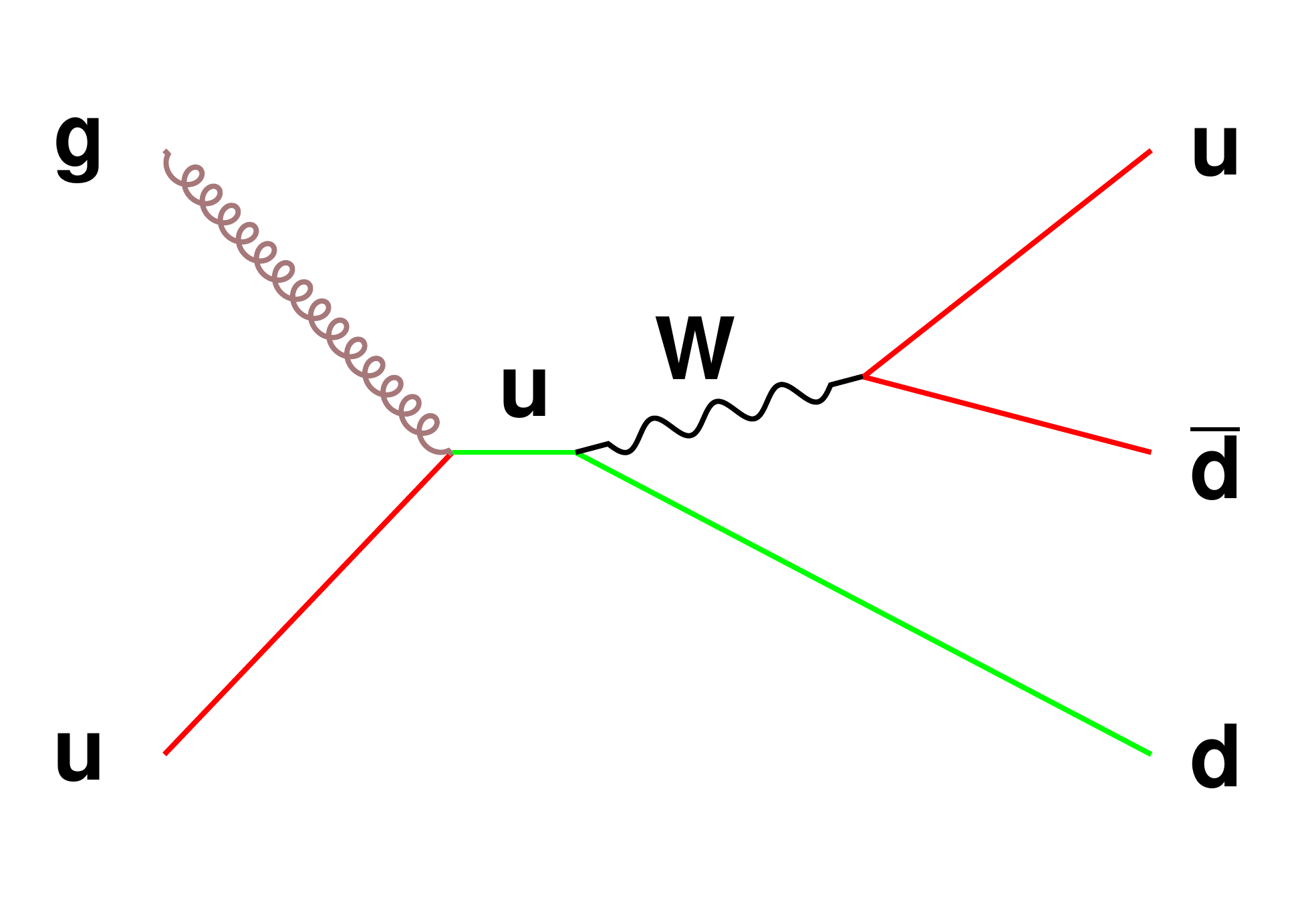}
 \includegraphics[width=3.8cm]{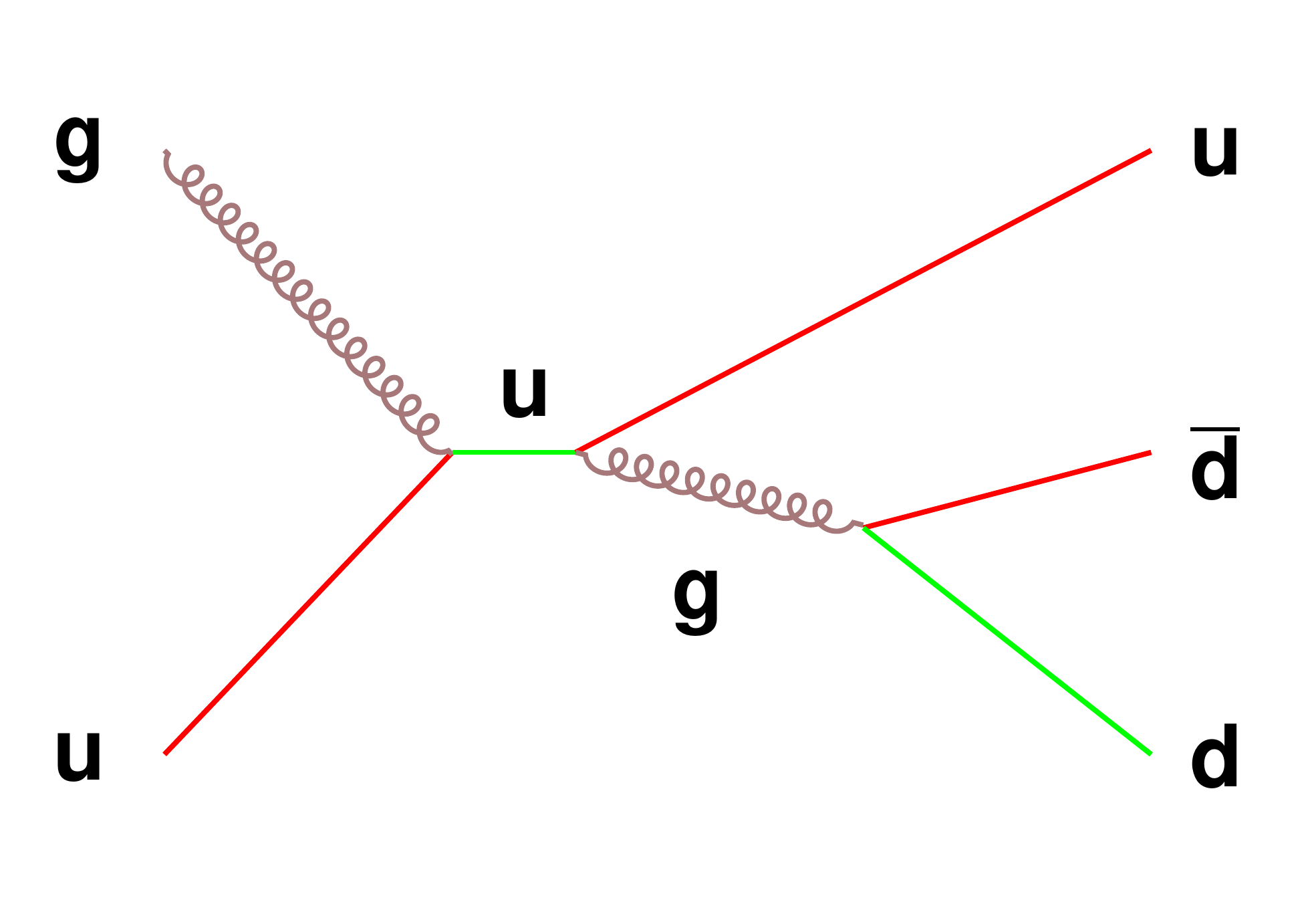}
 \includegraphics[width=3.8cm]{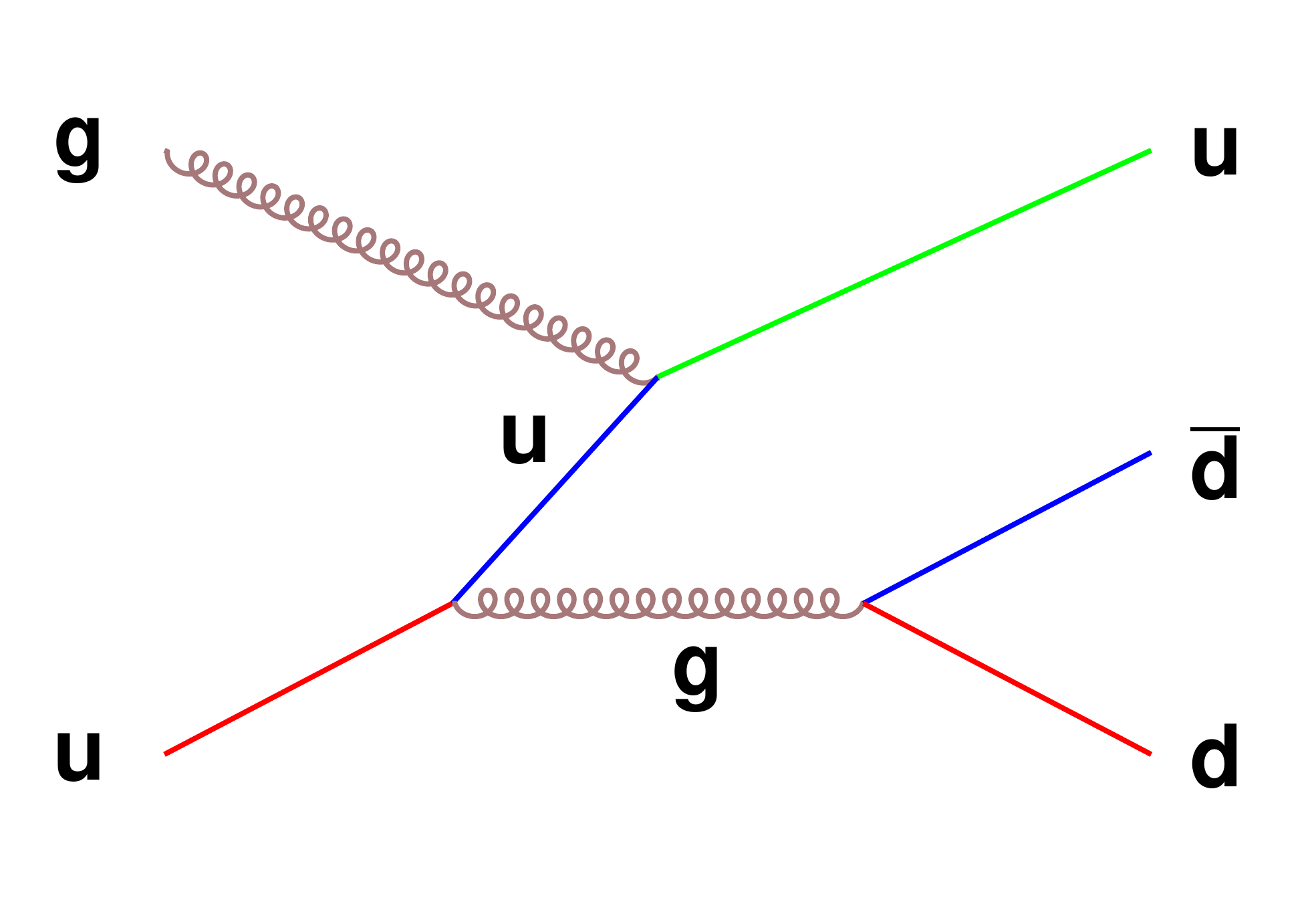}
 \vskip -0.2cm 
 \caption[]{Left: example boosted $W$ boson plus quark production diagram.
 Centre and Right: example QCD diagrams leading to the same
 quark state.}
 \label{fig:boostedW} 
 \end{figure}
 
Sample Feynman diagrams are shown in figure~\ref{fig:boostedW}.
It is an amusing feature of the QCD process in the centre that the
colour label of the internal $u$
quark is transferred first to the gluon and then to the final $d$
quark. Thus there is a $u$ and $\overline{d}$ pair that is a colour singlet in both
the electroweak and this QCD diagram. This does not, however, apply to the right-hand diagram as the incoming gluon changes the colour structure.

The cross-sections of the electroweak process ($W \rightarrow u\overline{d}$) are 0.11~pb for the $Wg$ process and 0.23~pb for
the $Wq$ process in a 2~\gevc2 window around the $W$ peak. The signal-to-background
ratios at the peak of the resonance are very large,  115
and 681 respectively. 
 This high signal-to-background ratio implies that interference effects must
 be small. Indeed, this is observed: the values of \etainf\ are
 1.004$\pm$0.002 for the $Wg$ process and 1.000$\pm$0.002 for the $Wq$ process, and the corresponding mass shifts are
 $-0.005\pm$0.004~GeV/c$^2$ and 0.007$\pm$0.004~GeV/c$^2$ respectively.

For the  $Wg$ process, the interfering QCD diagrams have the same
quark flavours in the initial and final state, much like the inclusive
production.
Therefore, interference effects will 
be smaller for $W \ra \cs$ than for $W \ra \ud$, in the same way as described for production at rest.
Conversely, for the $Wq$ process, only one of the final state quarks comes from the
initial state and as a result, the suppression will not be as strong. 
The  effects are so small because the $W$-like quark pair in QCD must
come from different vertices.

The analysis has been repeated with a
threshold of 200 GeV/c, where the interfering QCD component  is less suppressed. The signal-to-background ratios fall by a factor of four and the resulting peak shift
for $Wg \ra \ud$ is 0.011$\pm$0.005~GeV/c$^2$, and for $Wq$ it is $-0.001\pm0$.004~GeV/c$^2$. 
Thus, even for a threshold far below that currently employed for an
unprescaled jet trigger, interference can be safely neglected.


   
   \subsection{Inclusive Boosted  $Z$}
   \label{boostedz}

The inclusive boosted $Z$ production has many features in common with the $W$,
with possible recoil on a gluon, quark or antiquark considered.
The $Z\ra\bb$ channel is studied here and it is also compared with $Z\ra\uu$,
which is expected to show larger interference.
The simulation is very similar to the $W$ above, except that the \qq\ pair of interest is either \bb\ or \uu.
Unlike the $W$ case, two identical quarks can be produced, and as the
Sherpa selector acts on {\em all} matching combinations, it is changed
to require  50~GeV/c$^2$ or more. If this had not been done, the
joint requirements  that one of the partons has a \pt\ above 400~GeV/c and
the upper mass cut on both \qq\ pairs would together  have eliminated like-quark combinations.

   
   \begin{figure}[htb]                  
      \centering                        
      \includegraphics[width=3.8cm]{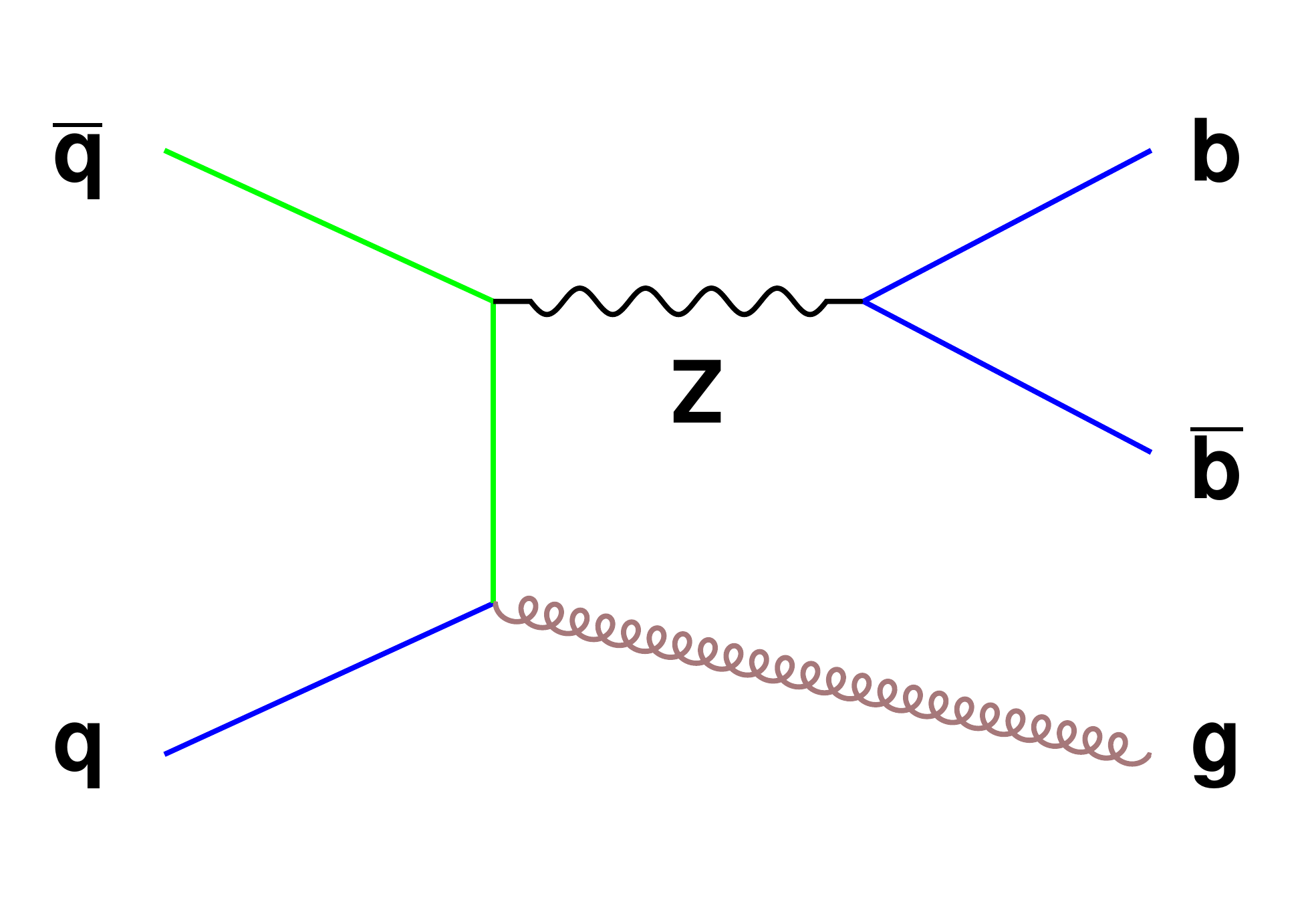}
      \includegraphics[width=3.8cm]{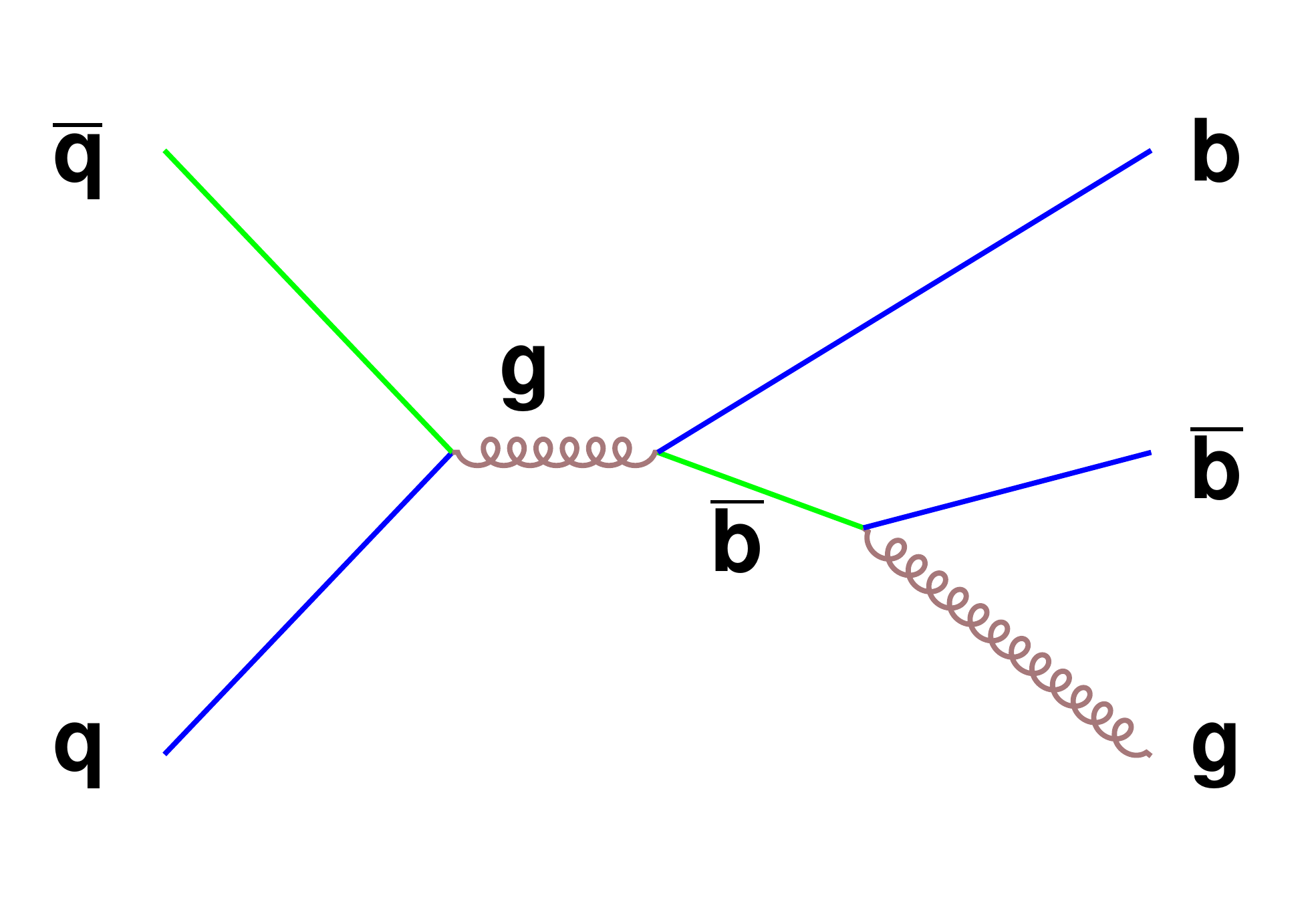}
      \includegraphics[width=3.8cm]{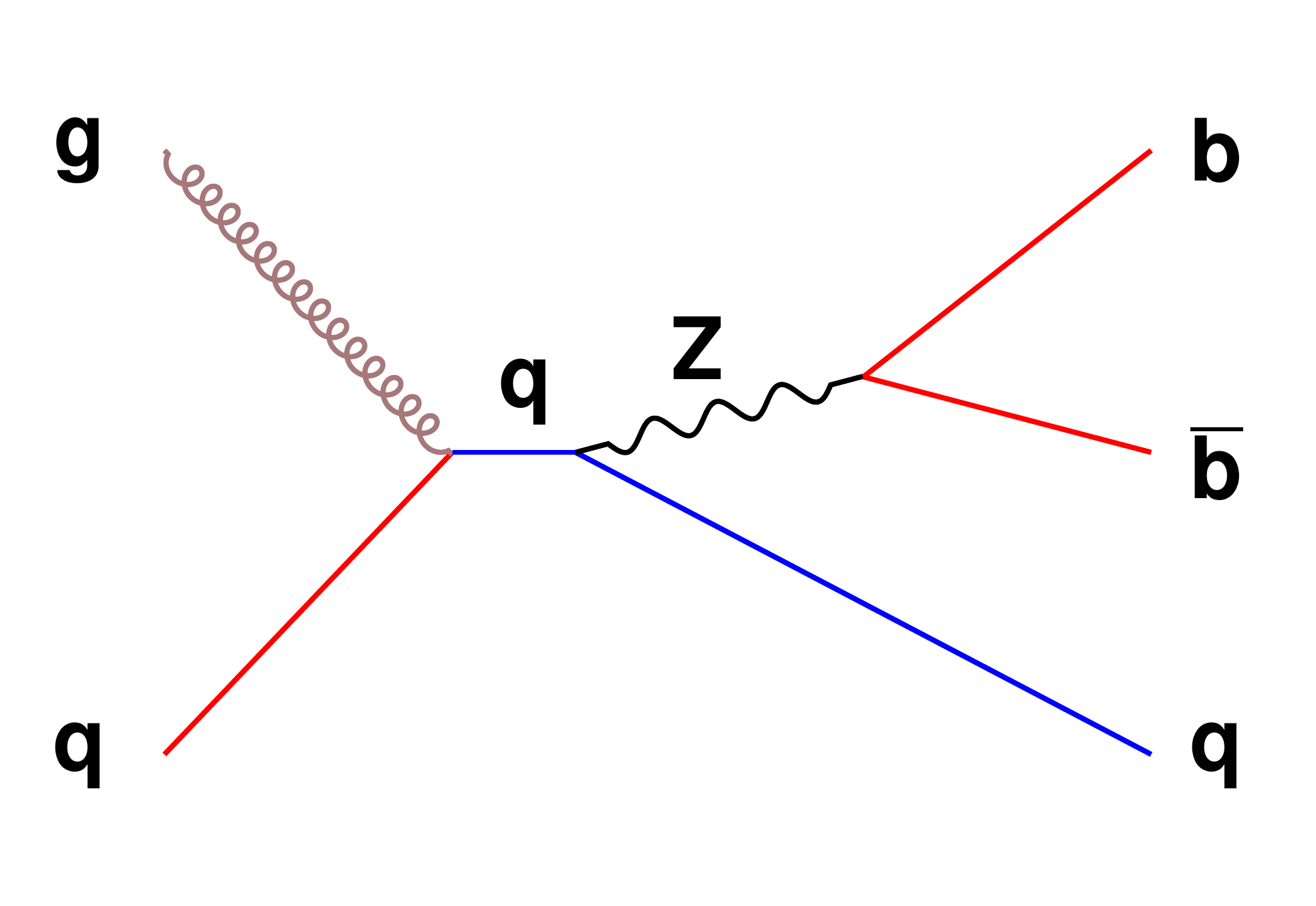}
      \includegraphics[width=3.8cm]{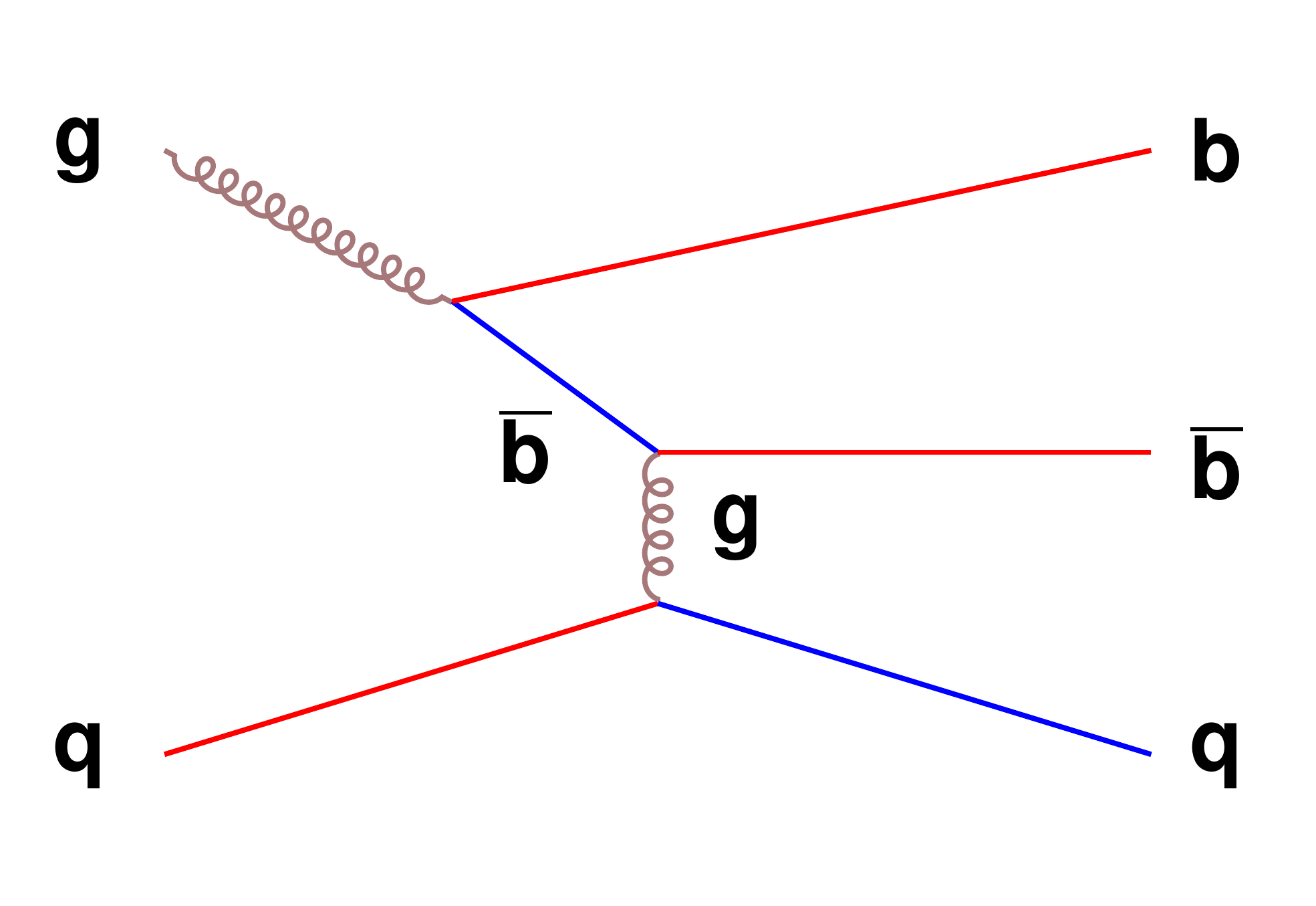}
      \vskip -0.2cm                     
      \caption[]{Feynman diagrams showing the production of a \qq\ pair
        recoiling against either a quark or a  gluon. The QCD
        examples  drawn are selected such that they can exhibit interference. }
      \label{fig:boostedZ}  
   \end{figure}

  Sample Feynman diagrams are shown in    figure~\ref{fig:boostedZ}.
  The rightmost diagram creates a colour singlet  \bb\ final state
  without requiring a $b$ quark in the initial state; there are other
  diagrams that produce a 
  colour octet \bb\  from gluon splitting. These are
   important differences from  the $W$ case. 
   

 The signal-to-background ratios and peak shifts are given in
 table~\ref{ta:boostedqg}, where results for the $W$ are also included. 
  The signal-to-background ratio is one to two orders of
 magnitude lower than for boosted $W$ production, so larger shifts
 might be seen, but most of this comes from gluon splitting. The largest shift is 0.022~\gevc2,
 which after applying a 10\% smearing produces a 0.18~\gevc2
 effect. This is negligible compared with the experimental
 systematic errors.

   \begin{table}[htp]
   \begin{center}
     \begin{tabular}{cc|lccr|lccr}
        Recoil             & \pt,                 & Boson       & signal /   & \etainf   & Peak shift       & Boson                        & signal /             &  \etainf     & Peak Shift \\
         particle          & GeV/c                & mode        & background &           & GeV/c$^2$        & mode                         & background           &        & GeV/c$^2$  \\
    \hline
    \multirow{5}{*}{gluon} & \multirow{2}{*}{400} & $Z \ra \uu$ & 9.3        & 1.004$\pm$0.002 & $+0.004\pm$0.005 & \multirow{2}{*}{$W \ra \ud$} & \multirow{2}{*}{116} & \multirow{2}{*}{1.000$\pm$0.002} &  \multirow{2}{*}{$-0.005\pm$0.004} \\
                           &                      & $Z \ra \bb$ & 21.        & 0.996$\pm$0.002 & $-0.009\pm$0.005 & & & \\
 \cline{2-10}
                           & \multirow{2}{*}{200} & $Z \ra \uu$ & 3.9        & 1.022$\pm$0.032 & $+0.022\pm$0.005 & \multirow{2}{*}{$W \ra \ud$} & \multirow{2}{*}{29}  & \multirow{2}{*}{0.996$\pm$0.002} & \multirow{2}{*}{$+0.011\pm$0.004} \\
                           &                      & $Z \ra \bb$ & 21.        & 0.996$\pm$0.002 & $+0.009\pm$0.004                                &                      & & \\
    \hline
    \multirow{3}{*}{quark} & \multirow{2}{*}{400} & $Z \ra \uu$ & 1.05       & 1.016$\pm$0.009 & $-0.004\pm$0.007 & \multirow{2}{*}{$W \ra \ud$} & \multirow{2}{*}{680} & \multirow{2}{*}{1.018$\pm$0.002} & \multirow{2}{*}{$0.007\pm$0.004} \\
                           &                      & $Z \ra \bb$ & 1.2        & 1.023$\pm$0.007 & $-0.002\pm$0.007 &                              &                      &           \\
 \cline{2-10}
                           &     200              & $Z \ra \bb$ & 0.91       & 0.958$\pm$0.008 & $+0.009\pm$0.004 & $W \ra \ud$                  & 160                  & 1.004$\pm$0.002 & $-0.001\pm$0.004 \\
  \hline
\end{tabular}
     \caption{Interference effects on the $Z$ and $W$ boson peaks with an initial state of either  $qg$ plus $\overline{q}g$ or $\qq$  and separated  by final state quark flavour. Signal-to-background is defined at the resonance peak without experimental resolution and using only the initial and final states quoted, with negligible statistical error.
   \label{ta:boostedqg}}
   \end{center}
   \end{table}

\begin{figure}[htb]                  
      \centering                        
      \includegraphics[width=\plotsize]{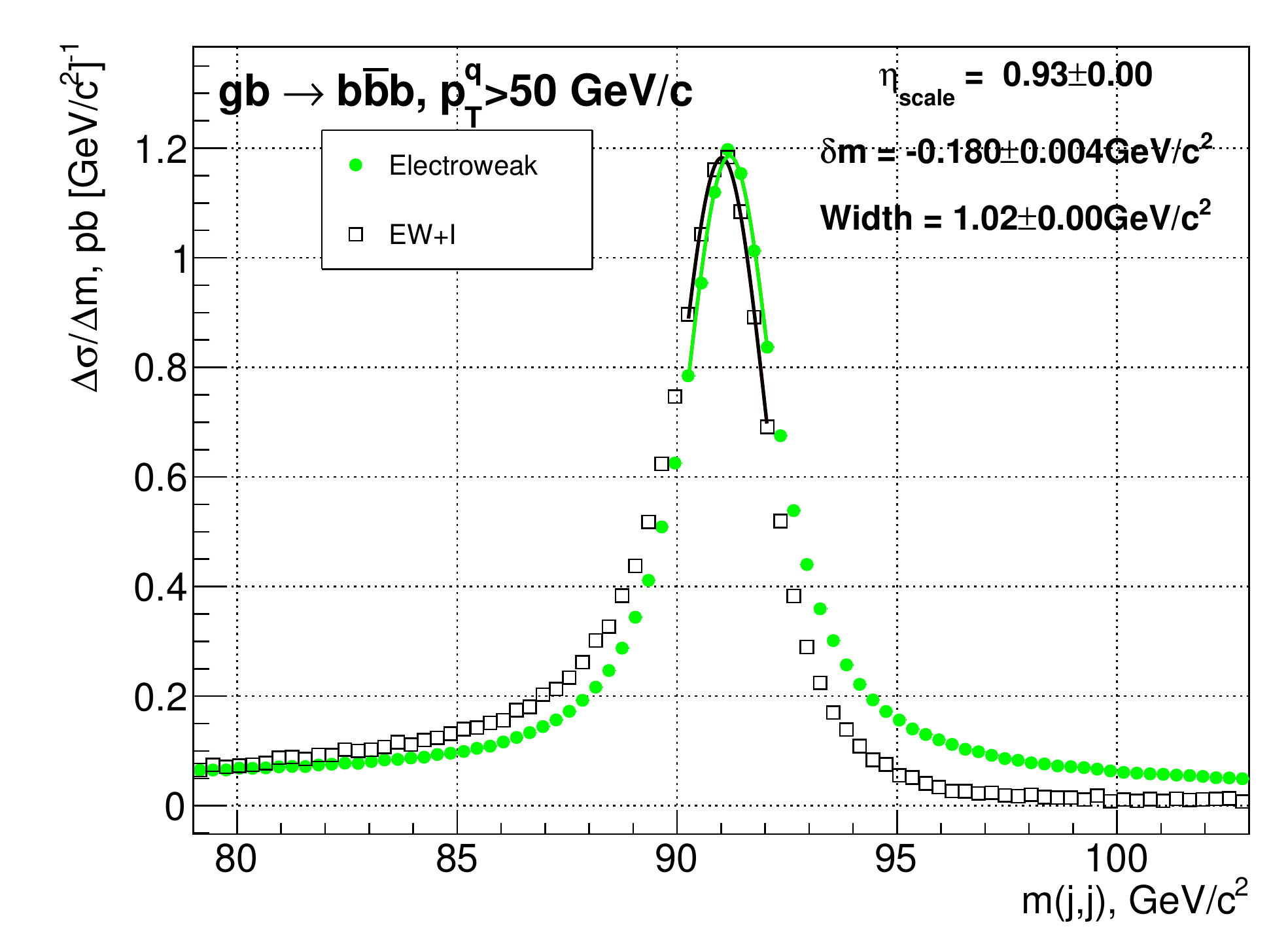}
      \includegraphics[width=\plotsize]{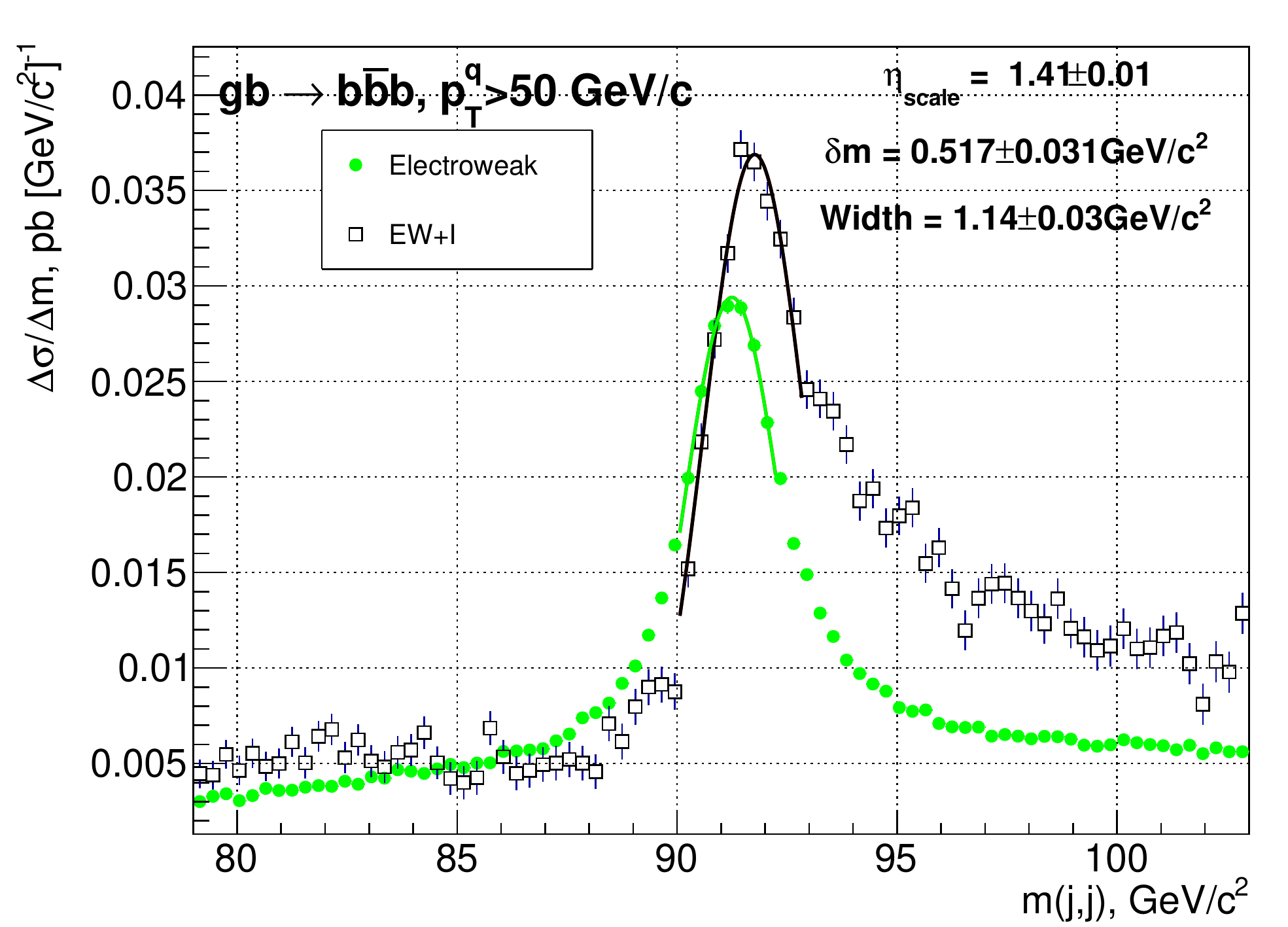}
      \vskip -0.2cm                     
      \caption[]{The process $bg  \rightarrow b \overline{b}b$. Left
        shows events where  $|\Delta\eta| <1$ 
        between the $Z$ candidate quarks, while right requires
        $|\Delta\eta| >2$. Each event generates two $\bb$
        combinations and  two entries in these plots if
        they both fall within the mass range.}
      \label{fig:zb-bb-50}  
   \end{figure}

   There is however one more thing which can potentially be measured
   in the $\bb q$ state - the flavour of the outgoing quark. If a
   third $b$ jet is identified, then diagrams like the one in
   figure~\ref{fig:boostedZ} (right) have enhanced possibilities for 
   interference, as there are two \bb\ combinations.
   Figure~\ref{fig:zb-bb-50} shows the electroweak and $EW+I$
   contributions when the minimum jet \pt\ requirement is reduced to
   50~GeV/c. 

  The \bb\ pairs  with similar pseudorapidity,  $|\Delta\eta| < 1$,
  show a significant shift of $-0.17$~\gevc2\ at parton level, which will
  correspond to a shift of more than a \gevc2 after detector effects are
  considered, which is similar to  energy scale systematic errors.
  However, those with a large $|\Delta\eta|$ separation show a dramatic effect,
  with  a shift in the peak position of around
  $+0.5$ \gevc2. This would produce a shift of several GeV in a measured
  spectrum.
  This effect  might be observable at the LHC. For example, the ATLAS
  collaboration in 2016~\cite{ATL-DAQ-PUB-2017-001} operated an
  unprescaled trigger 
  requiring two b-jets and a third jet with thresholds of 65, 65 and
  110~\gevc2. If an  analysis could be performed in the trigger system, or prescaled, 
  it is conceivable that those thresholds could be halved.

\section{$V$ + $\gamma$ production}
\label{sec:vgamma}
 
\begin{figure}[htb] 
 \centering 
 \includegraphics[width=3.8cm]{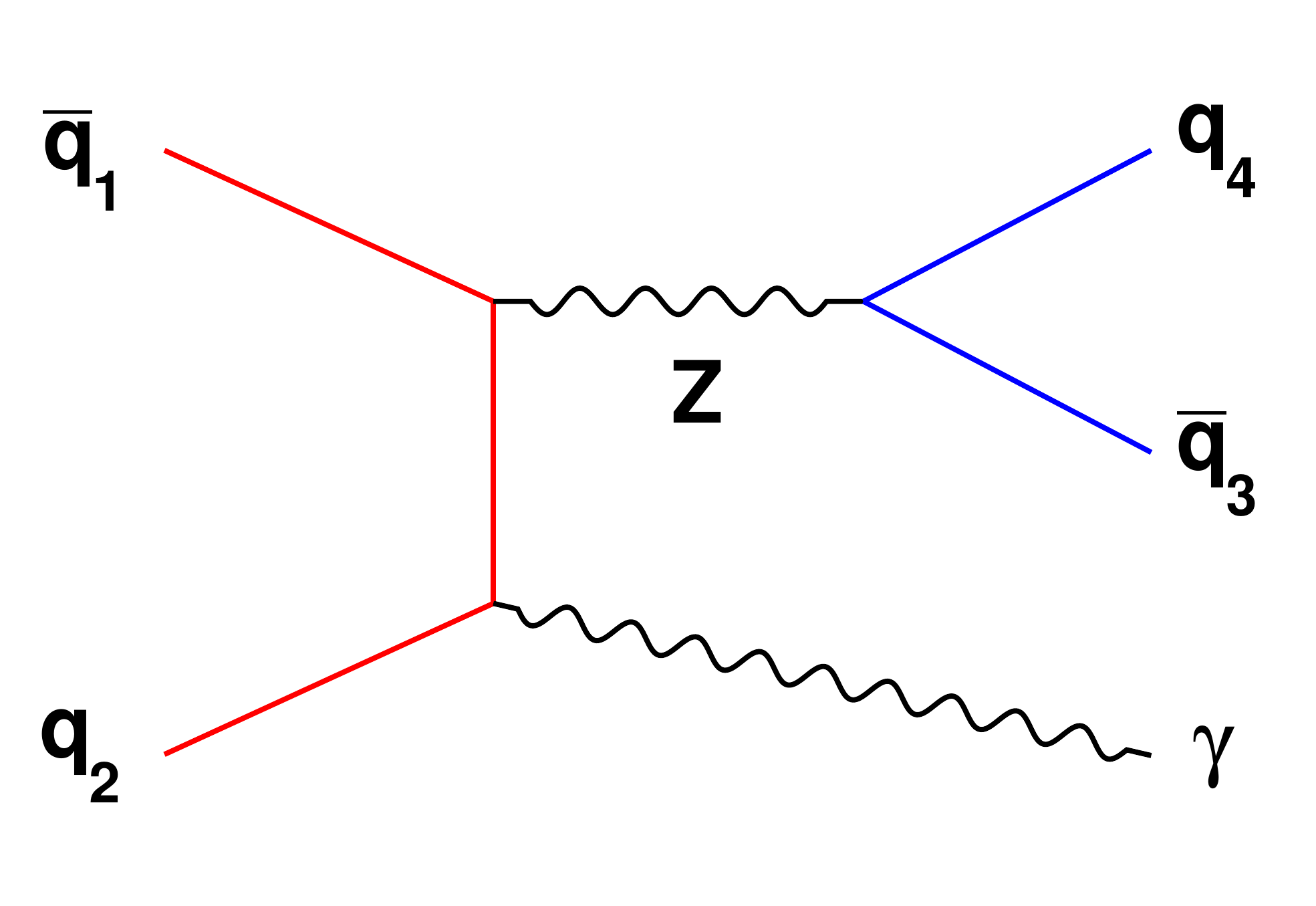}
 \includegraphics[width=3.8cm]{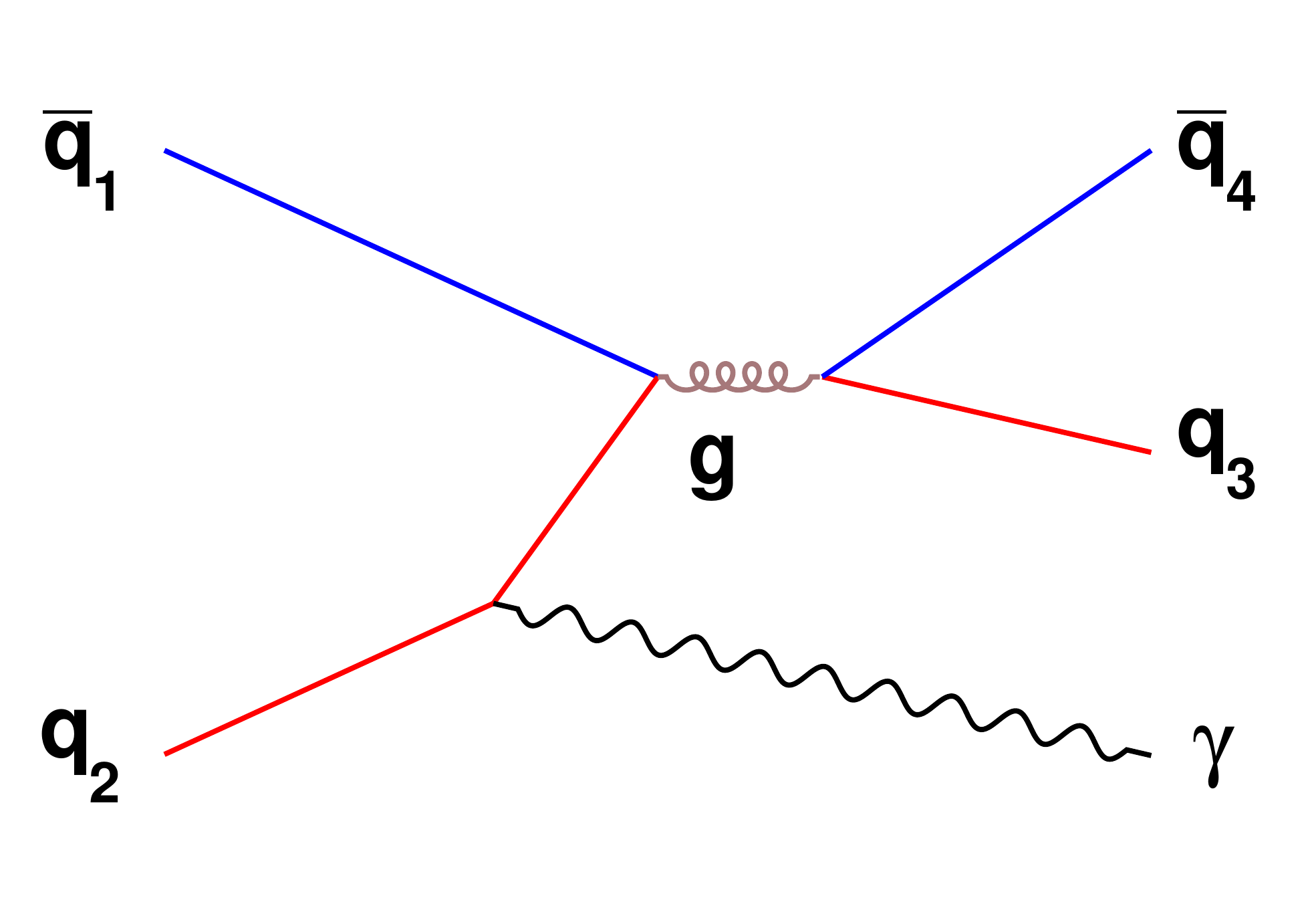}
 \includegraphics[width=3.8cm]{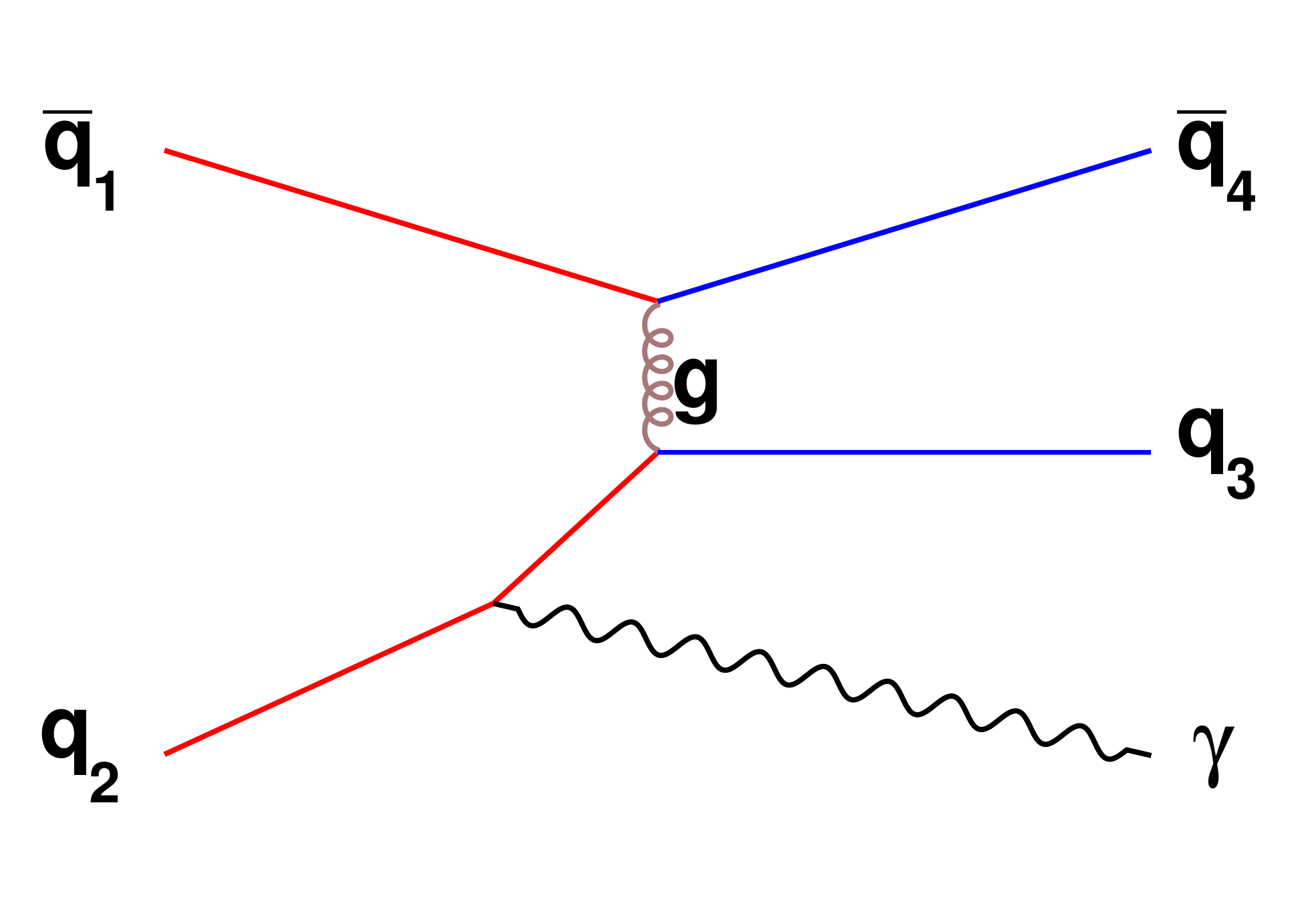}
 \vskip -0.2cm 
 \caption[]{Example Feynman diagrams for $\qq \ra \qq\gamma$. Left is an electroweak process
 and centre and right are referred to here as QCD.}
 \label{fig:boostedgam} 
 \end{figure}

Studying hadronic vector boson decays recoiling against a photon has significant experimental
advantages. Photons couple to charge, so the gluon background is
suppressed. ATLAS analysed $\gamma \bb$\cite{Aad:2019wdr}, with a
trigger on a single photon, an offline \pt\ selection of 175~GeV/c, 
and a $Z$ candidate \pt\ threshold of 200~GeV/c. For the current leading order
quark-level analysis, the \pt\ of the $V$ and the photon are exactly
equal, and 200~GeV/c is used as a benchmark for these processes. Sample 
Feynman diagrams are shown in
figure~\ref{fig:boostedgam}, and results are given in this section for
\uu, \bb, 
\ud\ and \cs\ quark combinations. As usual, the initial
state of $gg$, which does not interfere with the electroweak process at leading order, is not included.

 As always, 
 the colour structure of $s$-channel gluon production of the quark
 pair means it does not
 contribute to interference, although its rate is sizeable. The $t$-channel gluon exchange diagram can interfere, but it preserves the
 quark species, so while it is important for $\uu\gamma$, the small
 PDF suppress its contribution to the 
 \bb\ state.
 \begin{figure}[htb] 
 \centering 
 \includegraphics[width=\plotsize]{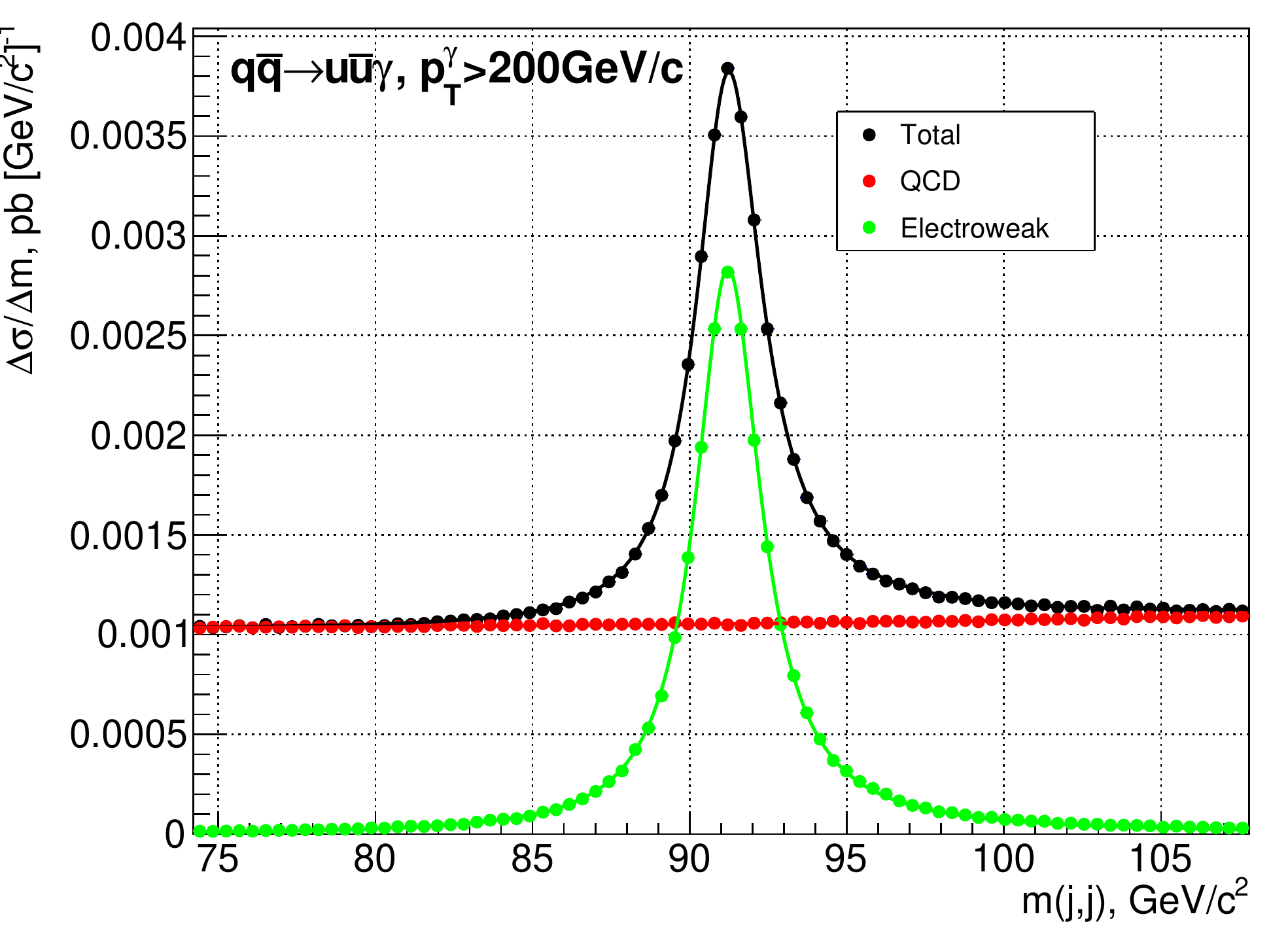}
 \includegraphics[width=\plotsize]{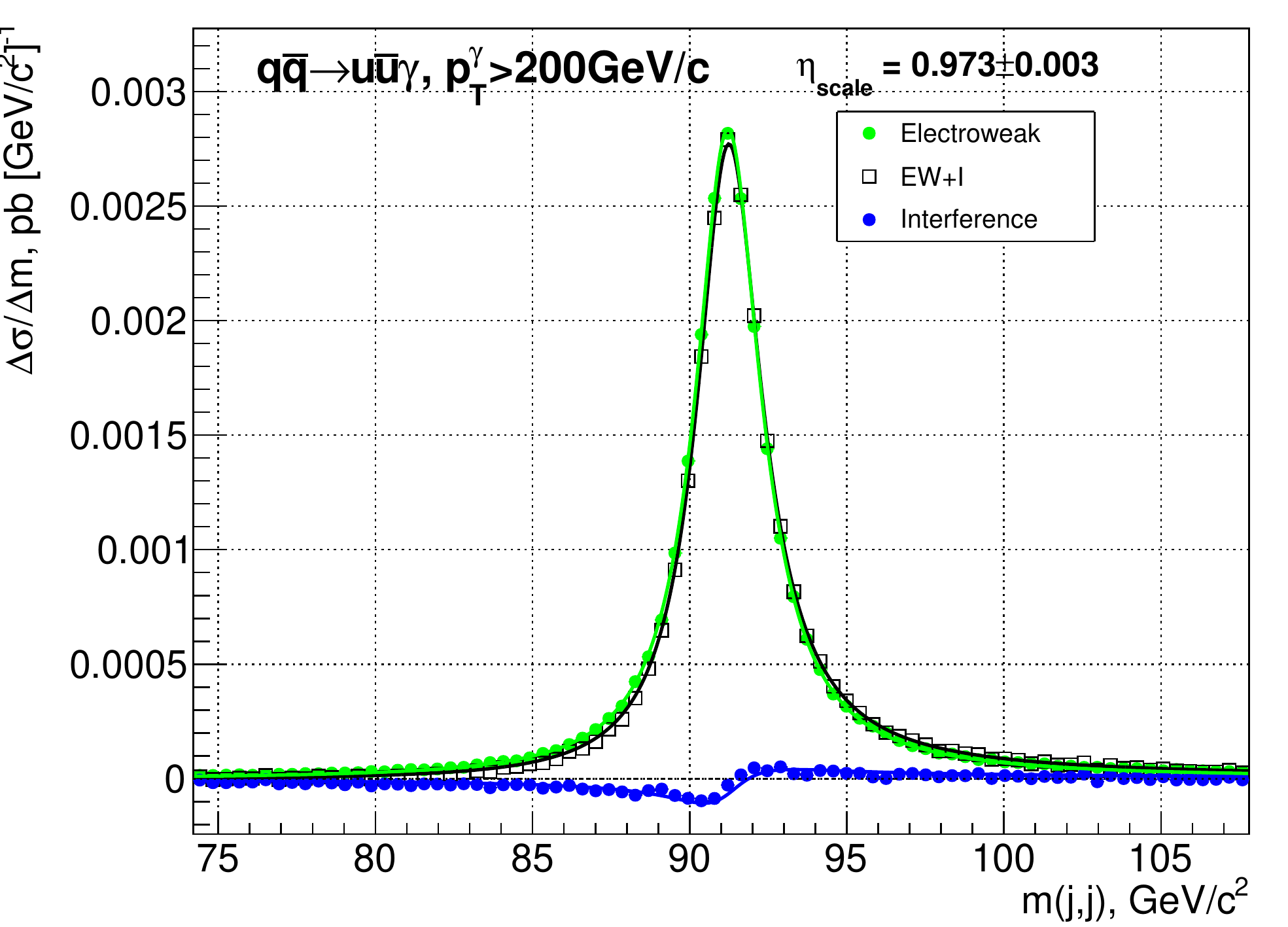}
 \vskip -0.2cm 
 \caption[]{The cross-section for the process $q \overline{q} \rightarrow u
 \overline{u} \gamma$, with the \uu\ pair required to have
 \pt$>$200 GeV/c. Left shows the pure electroweak (pale/green), pure
 QCD (dark/red) and total (black) cross-sections. Right has the same
 electroweak component, but the hollow black points show the total minus
 the QCD component, and the dark/blue is the interference term.}
 \label{fig:zgamma-uu} 
 \end{figure}
 
The case of $\qq \ra Z \rightarrow \uu$ is shown in
figure~\ref{fig:zgamma-uu}. The signal-to-background ratio, on
 peak, is nearly 3:1, but the  interfering fraction of the
 background, fitted as discussed in section~\ref{sec:inclusive},   is only $0.7\pm0.1$\%. This is consistent with  much of the background arising from 
 colour octet gluon splitting. The relative phase of the background is
 reversed in comparison to the inclusive case, giving a positive
 shift to the mass peak. The fitted shifts in the peak position are
 given in table~\ref{ta:boostedp}.

 \begin{table}[htp]
 \begin{center}
     \begin{tabular}{cc|lccr|lccr}
        Recoil             & \pt,                 & Boson       & signal /   & \etainf   & Peak shift       & Boson                        & signal /             &  \etainf     & Peak Shift \\
         particle          & GeV/c                & mode        & background &           & GeV/c$^2$        & mode                         & background           &        & GeV/c$^2$  \\
 \hline
 \multirow{8}{*}{photon} & \multirow{2}{*}{400} & $Z \ra \uu$ & 5.2 & $0.999\pm$0.002 & $+0.016\pm$0.005 \\
 & & $Z \ra \bb$ & 10.9 & $0.997\pm$0.002 & $-0.007\pm$0.004 \\
 \cline{2-10}
 & \multirow{2}{*}{200} & $Z \ra \uu$ & 2.7 & $0.973\pm$0.003 & $+0.045\pm$0.005 & \multirow{1}{*}{$W \ra \ud$} & \multirow{1}{*}{35} & $1.026\pm$0.002 & \multirow{1}{*}{$+0.007\pm$0.004} \\
 & & $Z \ra \bb$ & 9.7 & $1.002\pm$0.002 & $-0.008\pm$0.004 & \multirow{1}{*}{$W \ra \cs$} & \multirow{1}{*}{666} & $0.999\pm$0.002 & \multirow{1}{*}{$-0.001\pm$0.004} \\
 \cline{2-10}
 & \multirow{2}{*}{100} & $Z \ra \uu$ & 1.15 & $0.918\pm$0.005 & $+0.065\pm$0.007 \\
 & & $Z \ra \bb$ & 8.1 & $1.010\pm$0.002 & $+0.001\pm$0.004 \\
 \cline{2-10}
 & \multirow{2}{*}{ 50} & $Z \ra \uu$ & 0.46 & $0.769\pm$0.011 & $-0.073\pm$0.014 & \multirow{2}{*}{$W \ra \ud$} & \multirow{2}{*}{4.3} & \multirow{2}{*}{$0.953\pm$0.003} & \multirow{2}{*}{$-0.020\pm$0.005} \\
 & & $Z \ra \bb$ & 5.5 & $0.987\pm$0.002 & $-0.004\pm$0.004 & & & \\
 \hline
     \end{tabular}
          \caption{Interference effects on the  $Z$ and $W$ boson peaks in $\qq\ra V\gamma$,  separated  by final state quark flavour. Signal-to-background is defined at the resonance peak without experimental resolution and using only the initial and final states quoted, with negligible statistical error.
 \label{ta:boostedp}}
 \end{center}
 \end{table}


The shift is 0.045~\gevc2, which, when smeared by 10\%, moves
the peak position by 0.26$\pm$0.06~\gevc2. This is significantly
below the roughly 1\% experimental systematic errors on the mass scale. 
 The shift in $Z\ra\bb$ is only 0.008~\gevc2, far below any observable effects.

\begin{figure}[htb] 
 \centering 
 \includegraphics[width=\plotsize]{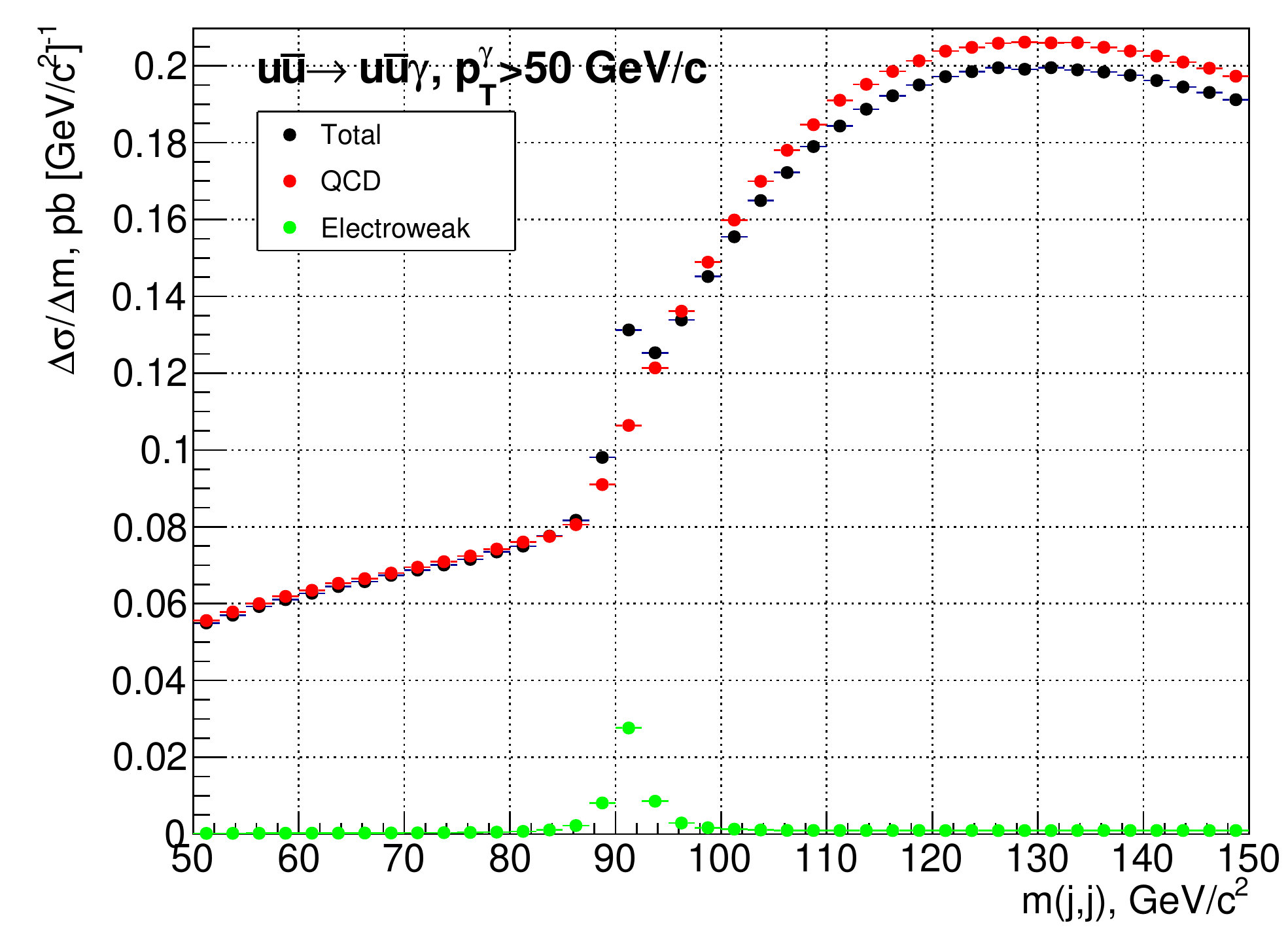}
 \includegraphics[width=\plotsize]{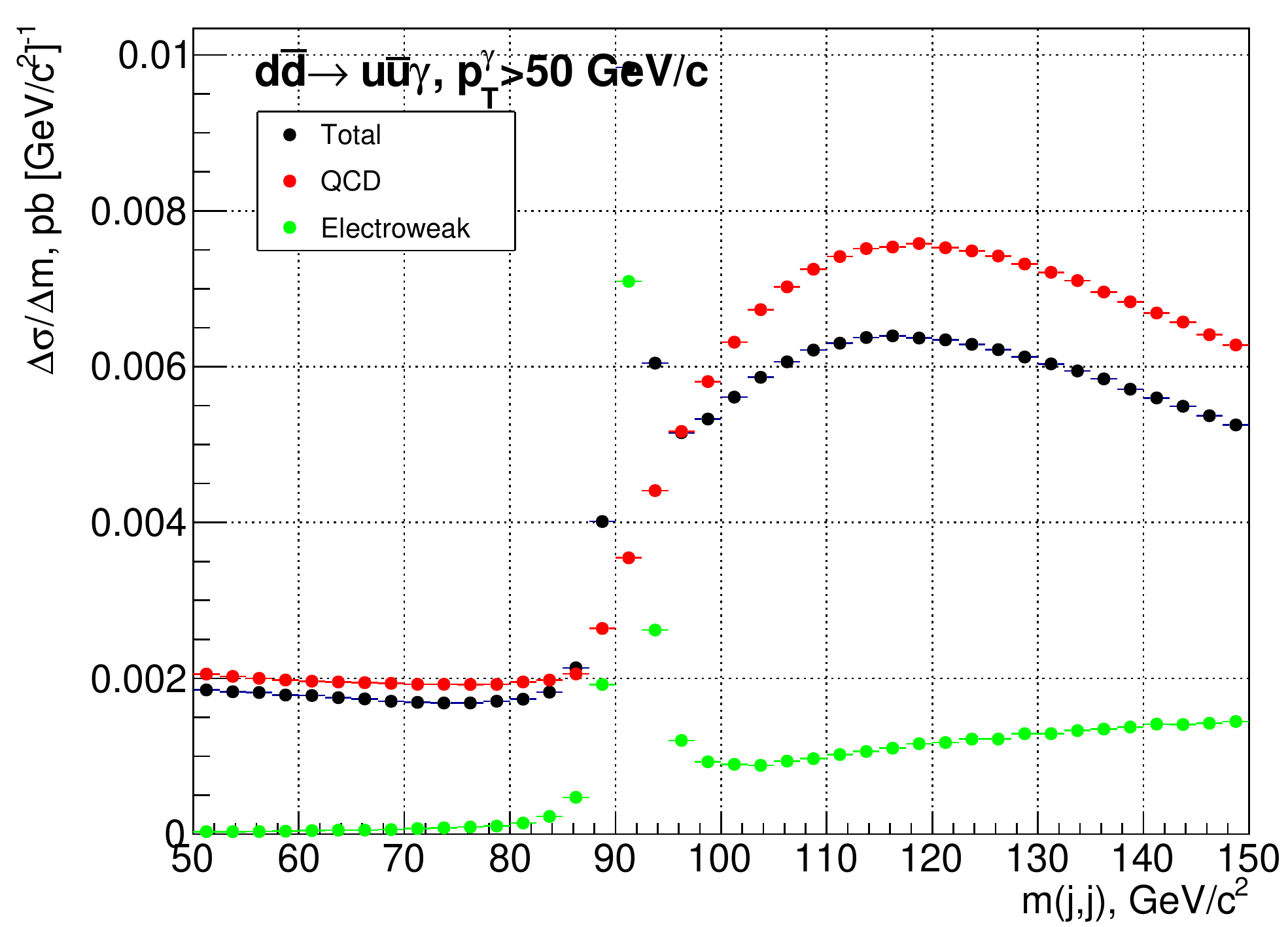}
 \vskip -0.1cm 
 \caption[]{The differential cross-section for the process $q \overline{q} \rightarrow u
 \overline{u} \gamma$, with the \uu\ pair required to have
 \pt$>$50 GeV/c. Left is $\uu \ra \uu \gamma$ and right is
 $\dd \ra \uu\gamma$. The differential cross-section of pure
 electroweak is pale/green, pure QCD is dark/red and the total is black. The electroweak
 $t$-channel component is much more prominent when $t$-channel
 $W$ exchange is possible, but in each case the 
 destructive interference is clear.}
 \label{fig:zgamma-uu-50} 
\end{figure}

A range of \pt\ selections is explored, from 400~GeV/c down to 50~GeV/c. The magnitude of the shifts grow as the \pt\ threshold is reduced. At
50~GeV/c   the $t$-channel electroweak 
exchange becomes important as evidenced by \etainf\ departing from 1. Figure~\ref{fig:zgamma-uu-50} shows the differential
cross-sections for the sub-processes $\uu \ra \uu \gamma$ and $\dd \ra \uu\gamma$. There is a qualitative  difference in the two processes which highlights the
role of the $t$-channel $W$ exchange that is only present in the $\dd \ra \uu\gamma$ process; however, both channels show clear interference. 
 With the 50~GeV/c \pt\ threshold requirement, the $t$-channel exchange components
 contribute mostly for 
 masses above the $Z$ mass, and so too does the
 interference, which suppresses the high-mass cross-section by
 approximately 20\%.
It must be stressed that experimentally these two processes are
indistinguishable.

Overall, the interference effects in the \uu\ channel are small, but in the
experimentally accessed \bb\ they
are close to zero. Again, the $b$ quarks PDF suppresses the
interfering background.

The $W+\gamma$ channel is probably experimentally accessible, although
there is no published study of the peak. However, as seen in
table~\ref{ta:boostedp}, the signal-to-background ratio is large, and the $s$-channel
interference is vanishingly small. In the same way as for the $Z+\gamma$ channel, at 50~GeV/c,
a $t$-channel electroweak component appears and complicates the picture.

\section{Vector boson associated production}
\label{sec:vv}

An interesting alternative source of hadronic vector bosons is diboson
production, with one of the bosons decaying leptonically. $WW$ and
$ZZ$ decays are explored here,  requiring that the leptonic decay
involves muons. 
Such events are 
triggered efficiently by the LHC detectors, right down to zero
recoil. 
All leptons, charged or neutral, are required to have a \pt\ above
25~GeV/c and the angular selections on the muons is $|\eta| <2.5$.

\begin{figure}[htb] 
 \centering 
 \includegraphics[width=3.8cm]{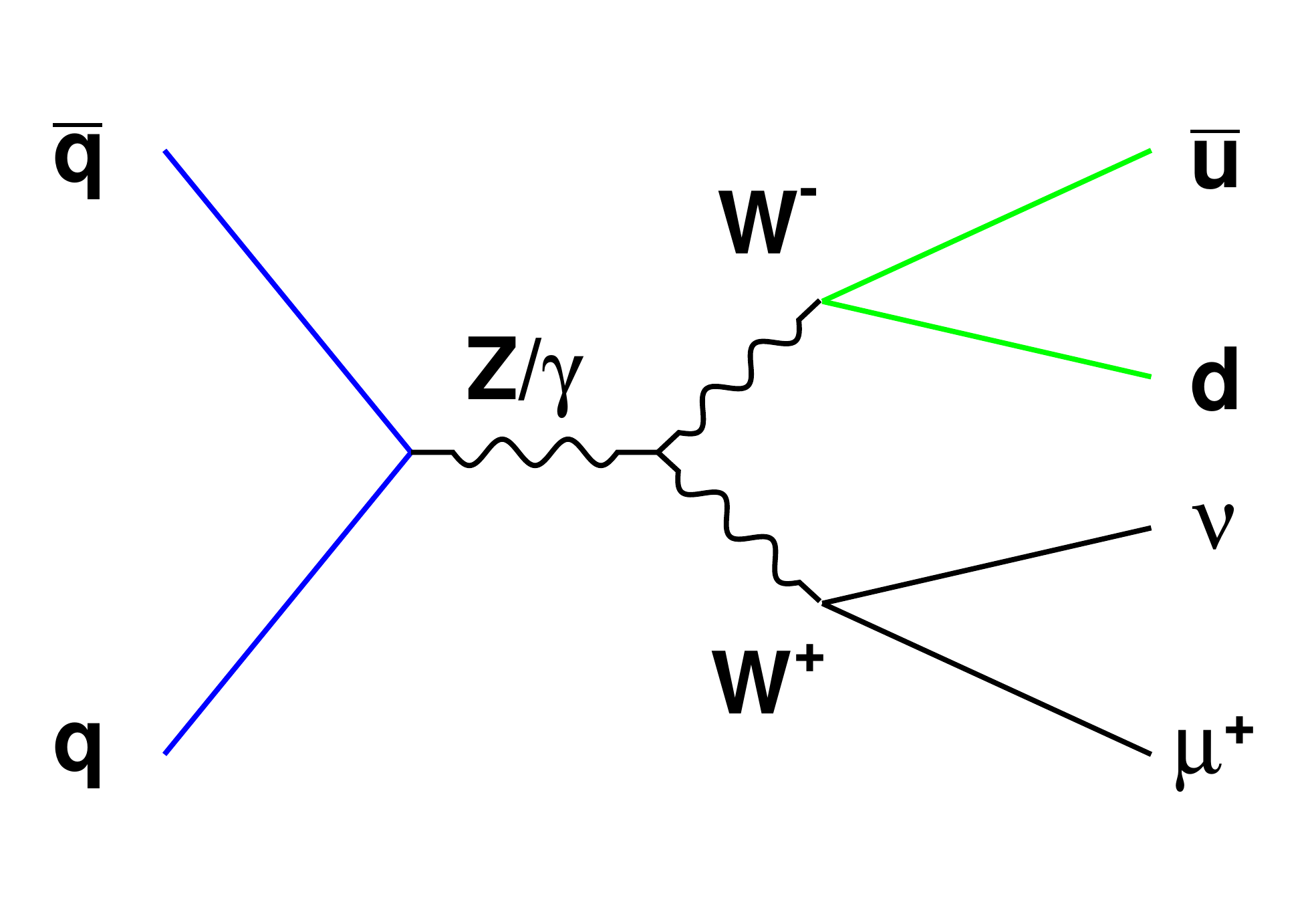}
 \includegraphics[width=3.8cm]{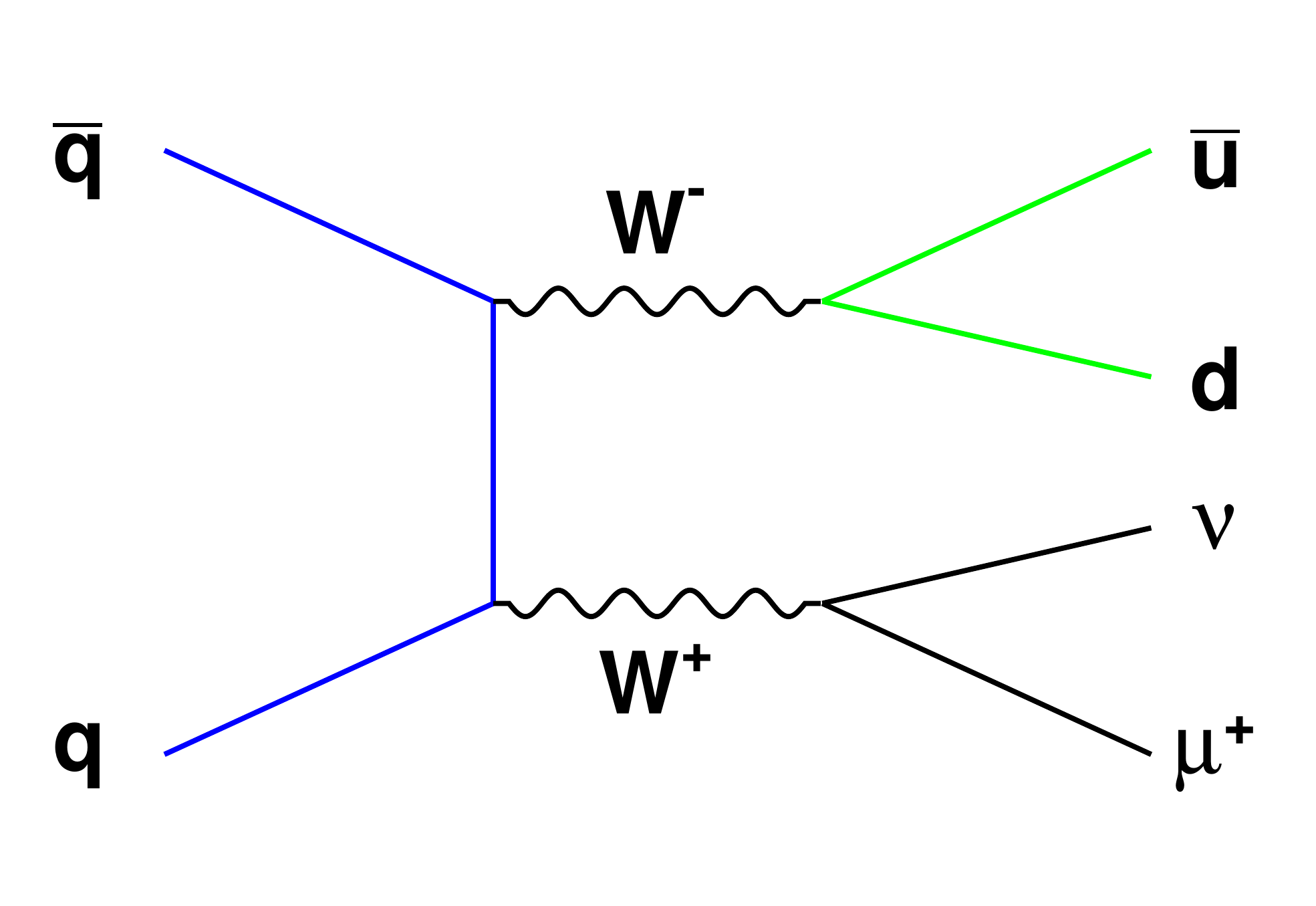}
 \includegraphics[width=3.8cm]{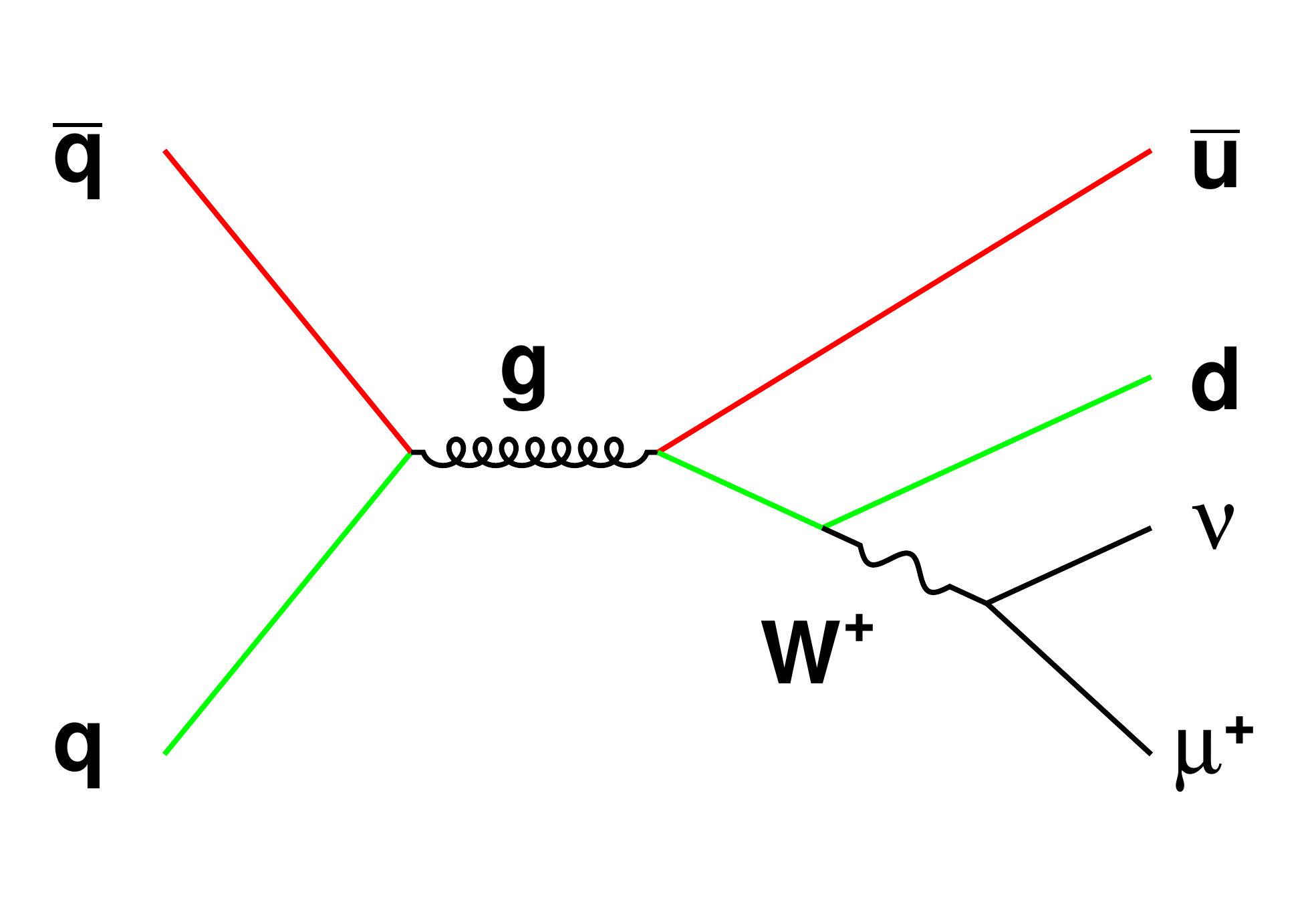}
 \includegraphics[width=3.8cm]{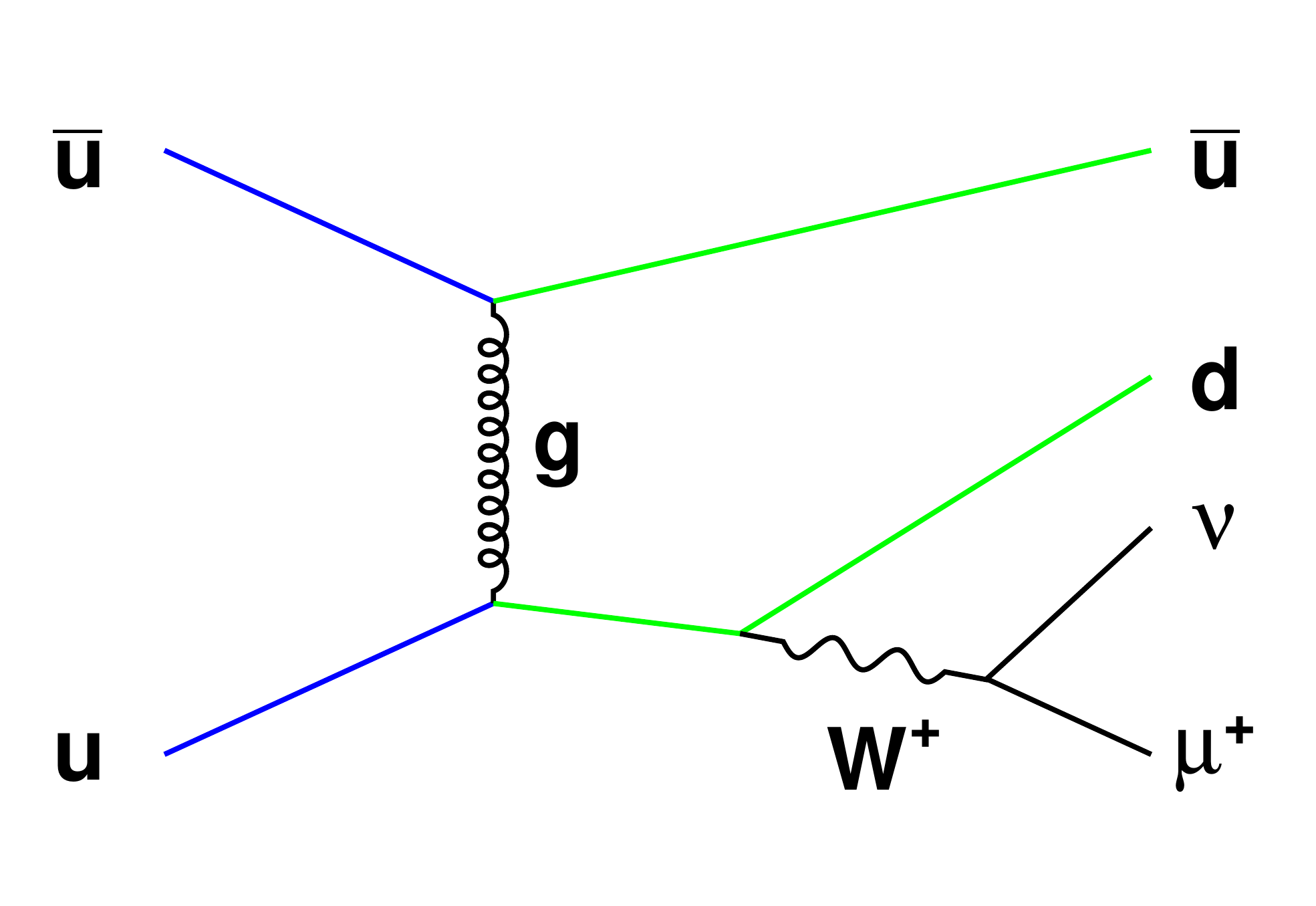}
 \vskip -0.2cm 
 \caption[]{The leading order \qq $\rightarrow$ $W^+$\ud\ diagrams. The first two diagrams show electroweak processes giving rise to on-shell $WW$ pair production,
 and the second two are the basic diagrams involving a gluon, referred to here as QCD .}
 \label{fig:lvqq} 
 \end{figure}

The background is similar to the previously-discussed modes with either \qq\ $s$-channel
interaction via a colour-octet gluon state, or a $t$-channel diagram
in which, for $WW$, the final state \du\ inherits the flavours of the initial
\uu\ system. The two diagrams are shown on the right-hand side
of figure~\ref{fig:lvqq} and, within the limitations of a diagonal
CKM-matrix, only \uu\ and 
\dd\ initial states can exhibit the $t$-channel process and thus interfere.
The signal has many  diagrams, involving  one $W^+$ along with
$W^-$, $Z$ or $\gamma$ contributions, but the on-shell process sought
has either a  triple-boson vertex or two separate $W$ boson emissions,
as shown in figure~\ref{fig:lvqq} (left).

\begin{figure}[htb] 
 \centering 
 \includegraphics[width=\plotsize]{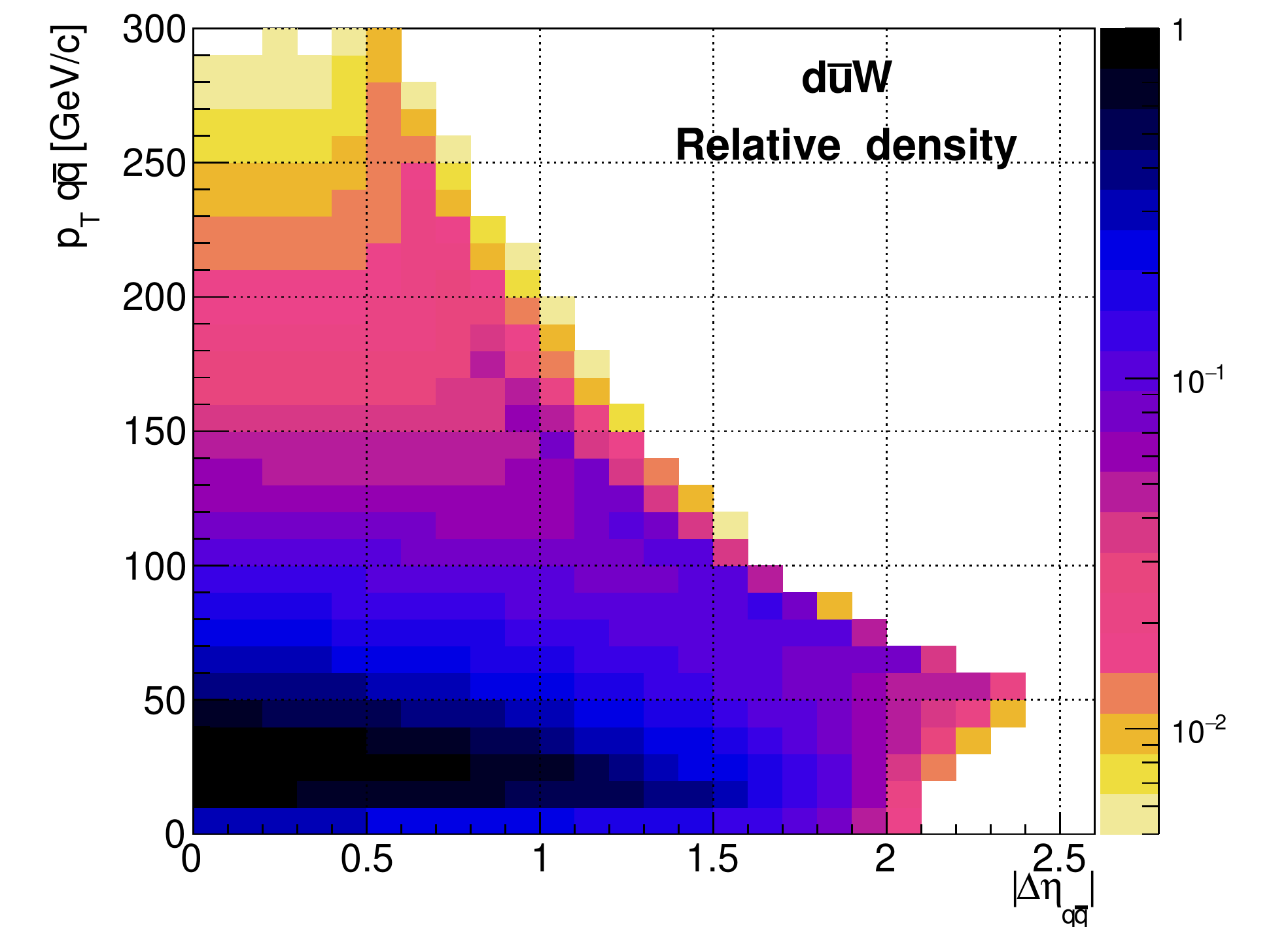}
 \includegraphics[width=\plotsize]{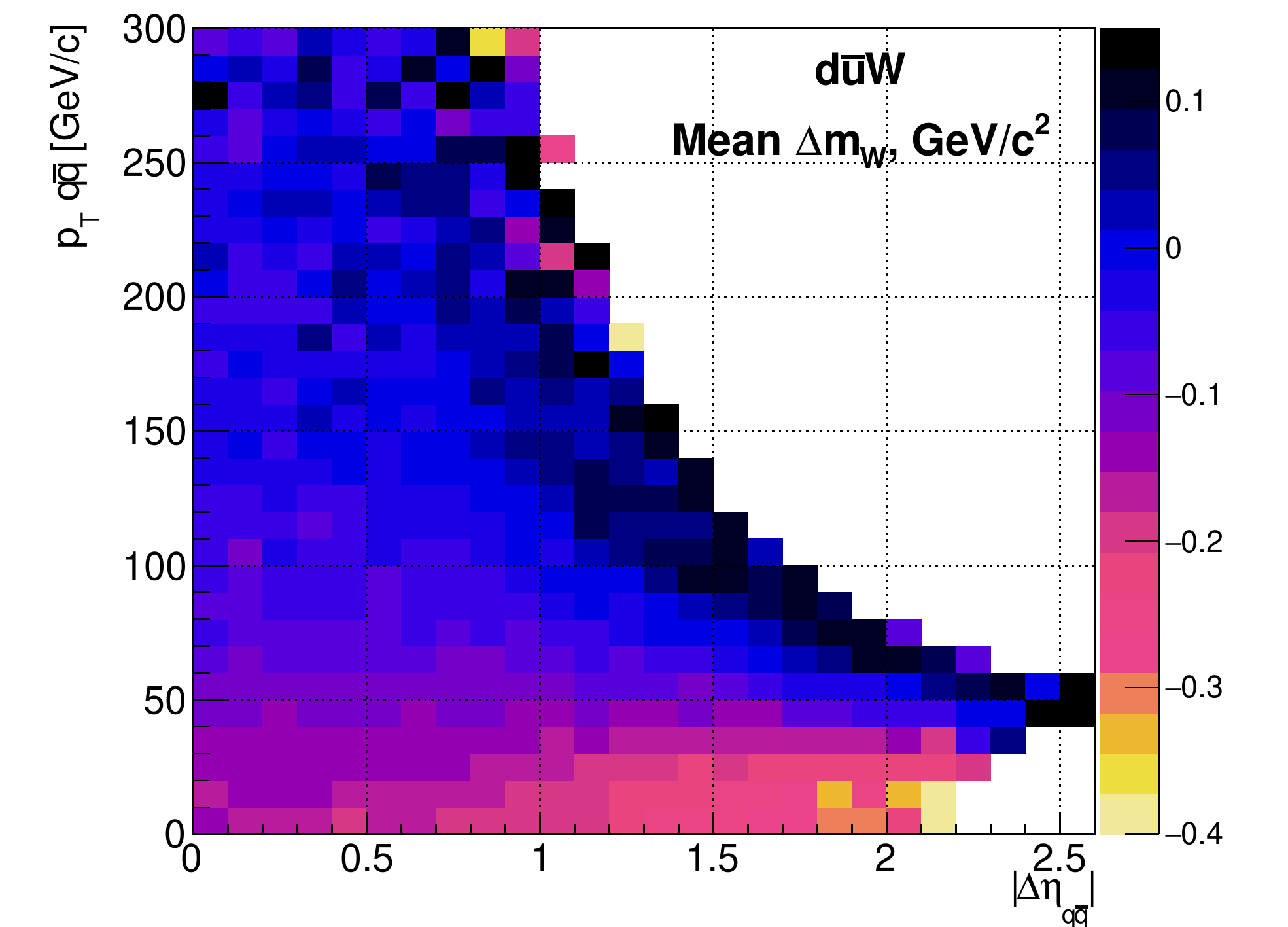}
 \vskip -0.2cm 
 \caption[]{The process $WW \rightarrow \mu^+ \nu \du$. Left
 is the relative density of the electroweak signal as a function of the
 $|\Delta\eta|$ and 
 \pt\ of the quark pair. Right is the shift in the mean mass due to
 interference, as defined in
 the text.}
 \label{fig:ww-2d} 
 \end{figure}

 The phase
 of the two electroweak diagrams shown is opposite, and their relative
 importance depends upon the kinematics, as does the QCD
 contribution. The inclusive shift in the peak position in
 $W^+$\ud\ is 
 $0.111\pm0.004$~\gevc2, which is significantly smaller than that
 seen in the prompt production. However, it is dependent on the
 event kinematics. This is highlighted in figure~\ref{fig:ww-2d},
 which shows (left) the relative phase space density and (right) the
 shift in the mean mass. The shift is calculated here by
 finding the mean mass of \du\ pairs within $\pm$2~\gevc2\ of
 80~\gevc2, and plotting the difference between the electroweak and  the
 difference between total and QCD as follows:

 \begin{equation}
 \Delta m = \frac{ \langle m_{total}\rangle\sigma_{total} -\langle m_{QCD}\rangle\sigma_{QCD}
   }{\sigma_{total}-\sigma_{QCD}} - \langle m_{EW}\rangle.
   \end{equation}

 This approach, calculating the  deviation from the mean in a window,
understates any  deviation from zero, but the bias has been found to
be small.

 The most probable kinematics has roughly the shift of
 $-0.1$~\gevc2 found for the mean, but when there is a
 pseudorapidity difference of 1.5 or more between the quarks the
 shifts range from 
 $-0.25$~\gevc2 at rest to $+0.1$~\gevc2 at recoil momenta approaching 100~GeV/c.
 Such effects, corresponding to two or three \gevc2 after allowing
 for detector resolution, are potentially observable. 

 \begin{table}[htp]
 \begin{center}
 \begin{tabular}{c|lccr|lccr}
  \pt,                & Boson & signal / & \etainf & Peak shift & Boson & signal / & \etainf & Peak Shift\\
 GeV/c                & mode & background & &  GeV/c$^2$ & mode & background & & GeV/c$^2$ \\
 \hline
 \multirow{2}{*}{100} & $ZZ \ra \uu$ & 6.9 & 0.981$\pm$0.002 & $-0.005\pm$0.005 & \multirow{2}{*}{$WW \ra \ud$} & \multirow{2}{*}{48} &\multirow{2}{*}{1.003$\pm$0.001} & \multirow{2}{*}{$+0.001\pm$0.003} \\
                      & $ZZ \ra \bb$ & 15.2 & 0.981$\pm$0.002 & $+0.007\pm$0.005 \\
 \hline
 \multirow{4}{*}{0}   & \multirow{2}{*}{$ZZ \ra \uu$} & \multirow{2}{*}{4.0} & \multirow{2}{*}{0.885$\pm$0.002} & \multirow{2}{*}{$-0.117\pm$0.005} & \multirow{1}{*}{$WW \ra \ud$} & \multirow{1}{*}{20.5} &\multirow{1}{*}{0.996$\pm$0.002} & \multirow{1}{*}{$-0.080\pm$0.005} \\
                      &       &             & & & \multirow{1}{*}{$WW \ra \du$} & \multirow{1}{*}{19.1} &\multirow{1}{*}{0.994$\pm$0.002}& \multirow{1}{*}{$-0.085\pm$0.002} \\
                      & \multirow{2}{*}{$ZZ \ra \bb$} & \multirow{2}{*}{13.1} & \multirow{2}{*}{1.006$\pm$0.002} & \multirow{2}{*}{$-0.010\pm$0.004} & \multirow{1}{*}{$WW \ra \cs$} & \multirow{1}{*}{86} &\multirow{1}{*}{1.011$\pm$0.002} & \multirow{1}{*}{$-0.027\pm$0.004} \\
                      &       &  & & & \multirow{1}{*}{$WW \ra \sc$} & \multirow{1}{*}{92} &\multirow{1}{*}{0.997$\pm$0.002}& \multirow{1}{*}{$-0.019\pm$0.004} \\
 \hline
 \end{tabular}
           \caption{Interference effects on the hadronic $Z$ and $W$ boson peaks in $\qq\ra VV$ with one $V$ decaying muonically and the other  separated  by final state quark flavour. Signal-to-background is defined at the hadronic resonance peak without experimental resolution and using only the initial and final states quoted, with negligible statistical error.
 \label{ta:boostedwz}}
 \end{center}
 \end{table}

 The mean shifts observed in a set of diboson states are summarised
 in table~\ref{ta:boostedwz}. $WW$ and $ZZ$ states are explored;
 $WZ$ are not expected to show qualitatively different behaviour. 

 
 The experimental study of such a state would not be able, in
 general, to distinguish jets from quarks or gluons. The  
 $W^+\ra \mu^+ \nu j j $  dijet  mass spectrum is analysed inclusively for the
 hadronic $W$ and $Z$ decays, and including all quark or gluon initial state combinations. The 10\%
 mass resolution is applied, and a mass region from
 70.775~\gevc2 to 92.525~\gevc2 is chosen, defined by the masses
 where the binned electroweak signal differential cross-section has halved
 compared with its peak. In that region, the pure electroweak cross-section
 is 1.3~pb, while that for QCD production of the additional jets
 is 20~pb. This corresponds to production of 180,000 events in
 140~fb$^{-1}$ of LHC data, or 700,000 if the 
 electron mode and their charge conjugates are also considered, which
 is may be  sufficient  to allow experimental study.

\section{Discussion}
\label{sec:discussion}

 The impact of electroweak-QCD interference on a variety of hadronic vector
 boson production modes has been explored. The effect is mostly to
 move  the peak position without changing the
 integral. There can be destructive interference
 effects between QCD and $t$-channel electroweak diagrams, for example
 in figure~\ref{fig:zgamma-uu-50}, but they are
 not the main focus of this study.

\begin{figure}[htb] 
 \centering 
 \includegraphics[width=\plotsize]{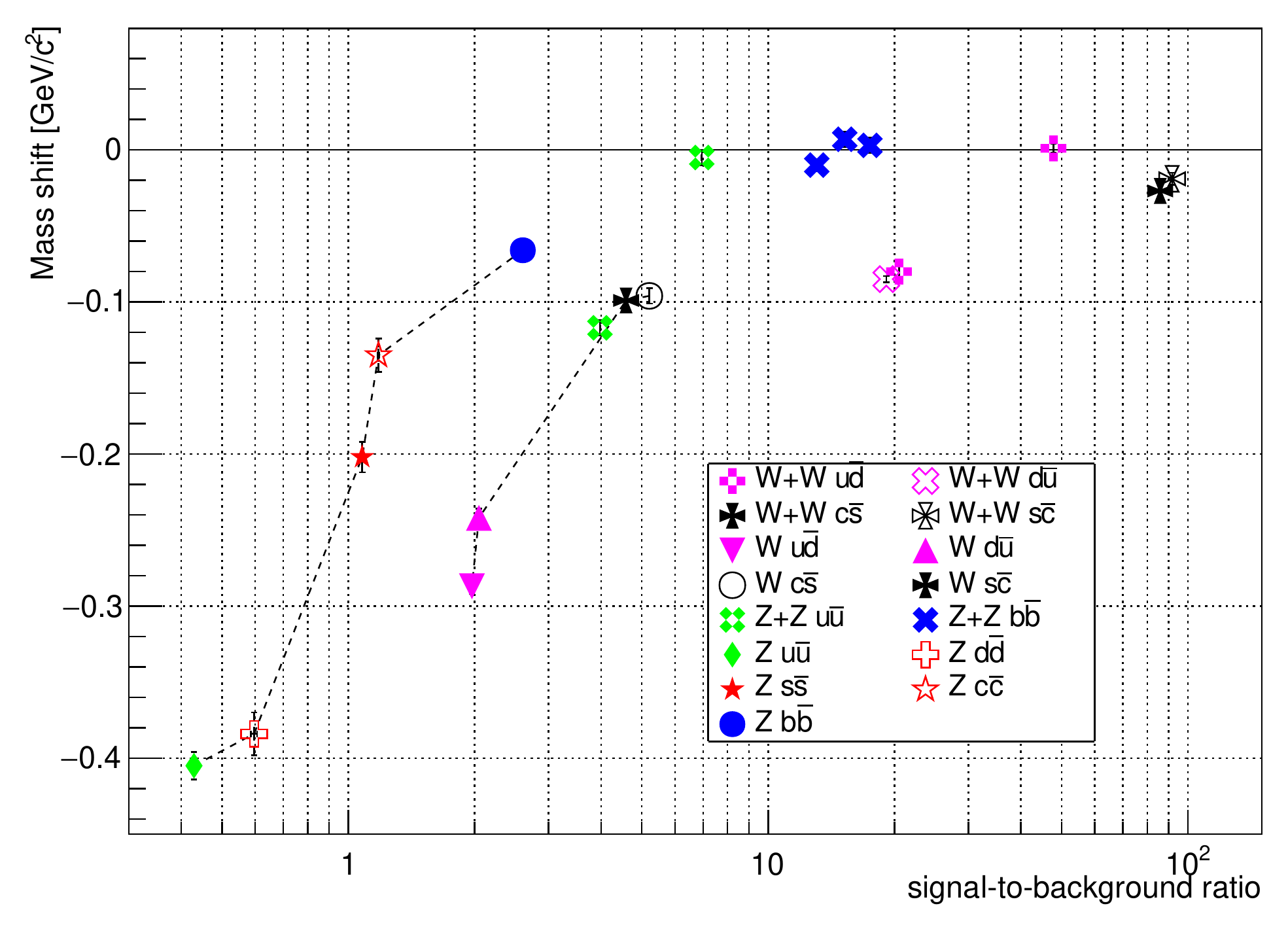}
 \includegraphics[width=\plotsize]{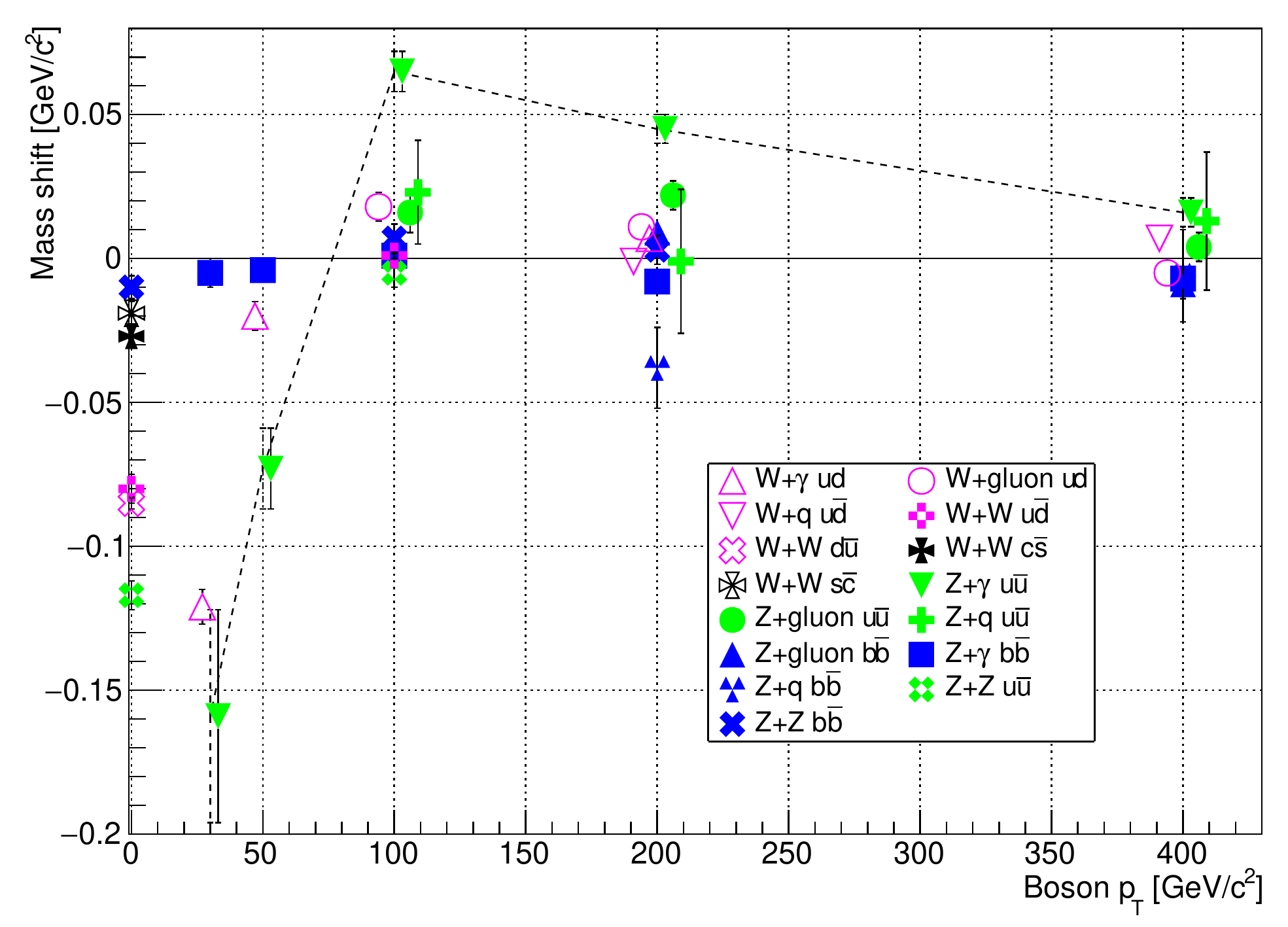}
 \vskip -0.2cm 
 \caption[]{Summary of the observed shifts in peak
 positions. 
 Left shows the peak
 shift as a function of the signal-to-background ratio in the final
 state, with inclusive production and recoil against a $W$ or
 $Z$.
 For both the $W$ and $Z$ produced singly, a dashed line links the
 different decay modes.
 Right shows hadronic vector boson recoiling against $q$,
 gluon, $\gamma$, $W$ or $Z$, as a function of the \pt\ of the
 vector boson.
 For  the  $Z\ra\uu$, a dashed line links the different \pt\ selections.

 }
 \label{fig:shift_summary} 
\end{figure}

 The mass shifts are summarised in
 figure~\ref{fig:shift_summary}, where the left-hand plot shows
 the production at
 rest, plotted against the signal-to-background ratio of the
 quark final state in question. The different decay modes of the  $W$
 and $Z$ decay at rest are linked to highlight the correlation between
 large
 signal-to-background ratio 
 and small  interference~\cite{Brooijmans:2018xbu}. Thus for the $Z$
 boson the interference is largest for \uu\ decay and smallest for \bb.
The $W$ boson does not have a gluon-splitting background, and thus
has higher signal-to-background than the $Z$, but the background is
more likely to have a colour singlet form, so the interference effects
are larger at a given signal-to-background. The peak shifts fall
inside the range spanned by the $Z$ decay modes.

Experiments generally uses $b$-tagging to identify $Z$ bosons. This
suppresses the $t$-channel QCD processes and reduces the interference
effects rather accidentally.
If there were a pure and efficient way to identify $u$-quarks
 available, experiments would use that to identify $Z$ candidates
 too, and  would then experience larger interference effects.
 
 It  may be possible to study inclusive $Z\ra\bb$  using LHC data. The ATLAS and CMS collaborations record data at about 1~KHz, from an LHC bunch-crossing rate of around 30 MHz. Thus one event in every 30,000 is recorded as pile-up in the detectors, or approximately 4~pb$^{-1}$. If a 10\% resolution is assumed the  $Z\ra\bb$ cross-section in $80-100$~\gevc2 at leading order is  2500~pb, and the QCD background in \bb\ is only 8000~pb. If non-$b$ background can be suppressed using  $b$-tagging, this should be clearly visible.

In the light of the dependence on the signal-to-background ratio in \ref{fig:shift_summary} (left), an explorative simulation
of $t\overline{t}$ has been performed. This required the process 
$gg \rightarrow \mu^- \overline{\nu_\mu}\overline{b} \du b$, with
selections on the two
$W$ candidates masses between $50$~\gevc2 and $120$~\gevc2, and the top candidates masses between 160~\gevc2 and 190~\gevc2. The cross-section with two electroweak vertices, one 
$W$ boson plus QCD jets, is 0.26$\pm$0.01~fb, while that at fourth
order in electroweak, which includes two on-shell top quarks, is
3.01$\pm0.05$~pb. This signal-to-background ratio, 10$^4$, strongly 
suggests that interference will have a completely negligible impact.

The mass shifts calculated in  various boosted states are shown against the
 vector boson minimum  \pt\ requirement in
 figure~\ref{fig:shift_summary} (right).
As with the bosons at rest, they are largest for $Z\ra\uu$, and
smallest for $Z\ra\bb$, with the  $W$  bosons being
intermediate.
 When vector bosons are boosted to 200~GeV/c or more, there
is generally a  positive shift, but it is too small to be
experimentally observable.
For transverse momenta of 50 to 100~GeV/c the patterns are complicated,
with the interference swapping sign, partly because the electroweak process is
no longer dominated by $s$-channel resonances. The 2D  space of $|\Delta\eta$
between the quarks and the vector boson \pt\ is presented   in
figure~\ref{fig:ww-2d} for the $WW$ case, where the varying contributions of
TGC production and emission of two independent $W$ bosons, which have
opposite phase,  leads to opposite sign shifts in different regions.

The shifts in the peak positions are increased by roughly an order of
magnitude when smearing the results  to mimic resolution
effects. For example, an intrinsic shift of $-0.409\pm0.005$~\gevc2 in $Z\ra\uu$\ is
enlarged to
$-3.4\pm0.2$~\gevc2.
This is reproduced by an approximate 
$s$-channel-only interference formula. In this context, the parton shower
simulation acts similarly to detector resolution. This is not
surprising as it is independently simulated and knows nothing of the
interference in the matrix elements. Simulation using Delphes
suggests that slightly smaller effects would occur as the jets with large
$|\Delta \eta|$ are less likely to be accepted. This could be offset
by working with a lower jet \pt\ threshold than the 25~GeV/c generally
used here.

For the original question that considers whether the boosted hadronic bosons are good
standard candles for the calibration of the detectors, the answer is
yes. The $W$ decays, especially boosted, have excellent intrinsic
signal-to-background ratio, while the $Z\ra \bb$ decay mode is protected by
the colour structure of QCD and the small $b$
PDF. The impact of interference on kinematics close to the published
measurements has always been found to be small. The one caveat is that
NLO effects might give rise to colour-singlet \qq\ pairs with a
significant cross-section, which would reopen the question for the
$Z$ boson.
There is no analogous QCD
process for \ud\ production, so the $W$ boson results should be
regarded as more robust.
Higgs to \bb\ decay has a cross-section significantly lower than $Z$,
but its 4 MeV width improves the signal-to-background ratio by 3 orders of
magnitude and interference can have no practical impact there.

In future trigger-level analyses, for example in the $W\gamma$
state with a 50~GeV/c \pt\ threshold, the $s$- and $t$-channel interference effects might perhaps be
possible to observe. More challenging to record is the $b\bb$ final state, where large effects should be
observable if significant luminosity can be recorded with a
\pt\ threshold of  50 GeV/c. This $Z$ mode  might be enhanced by NLO
QCD corrections. 

However, the $WW\ra l \nu \qq$  process seems to have the greatest promise for
experimental study. The $W\ra\du$ mass peak, after detector resolution,  is
expected to change position by two or more \gevc2 across the kinematic
plane. Smaller effects will occur for \cs, but it might be possible
to distinguish these final states using charm tagging.
When only irreducible background is considered, and pileup neglected,
the signal-to-background ratio  
exceeds 5\%, so a more detailed study seems warranted.
This particular example includes a leptonically
decaying boson, so the experiments have recorded this data,  and the
experimental collaborations are encouraged to see if the predicted effects can
be observed.

\begin{acknowledgements}
Great thanks are due to  J. Quevillon who clarified many of these
concepts for us and suggested studies and cross-checks, L. Xia, for
the helpful observation that at NLO colour octet and singlet merge and to S. Schramm, for the observation that pileup events contribute a significant trigger-free event rate. 
We also wish to thank our journal referees, whose insistence that we
explore the impact of the jet \pt\ threshold and the use of Delphes
simulation highlighted important issues.
\end{acknowledgements}

\bibliographystyle{spphys}      

\bibliography{epjc}

\appendix

   \section{Example Run.dat}
   \label{sec:run}

   An example `run.dat', for $\mu^+ \nu \du$ production with a
   \pt\ selection set to 0 GeV/c in the second last line.
   
\begin{verbatim}
(run){
  # general settings
  EVENT_OUTPUT=HepMC_GenEvent[sb]
  EVENTS 25;

  # me generator setup
  ME_SIGNAL_GENERATOR Comix;
  SCALES VAR{Abs2(p[0]+p[1])};
  
  # Five lines added to switch off decays.
  SHOWER_GENERATOR=None
  FRAGMENTATION=Off
  MI_HANDLER=None
  ME_QED=Off
  BEAM_REMNANTS=0

  # LHC beam setup:
  BEAM_1 2212; BEAM_ENERGY_1 6500;
  BEAM_2 2212; BEAM_ENERGY_2 6500;
}(run)

(processes){
Process 94 94 -> 1 -2 -13 14  ;
  Order (*,4);
  Print_Graphs graphs;
End process;
}(processes)

(selector){
	Mass 1 -2  50 150
	Mass -13 14  70 90
	PT 93 25 E_CMS
	PT 90 25  E_CMS
	PseudoRapidity 93 -2.5 2.5
	"Calc(PPerp(p[0]+p[1])>0)" 13,-14 1,1
}(selector)
\end{verbatim}

\end{document}